\documentclass[iopjournal,nofootinbib]{revtex4-2}

\usepackage{amsmath}
\usepackage{amsfonts}
\usepackage{amssymb}
\usepackage{braket}
\usepackage{tcolorbox}
\usepackage{xcolor}
\usepackage{graphicx}
\usepackage{enumerate}
\usepackage{subfigure}
\usepackage{hyperref}

\begin{document}

\title{Emulating XX catalysts for quantum annealing via self-consistent transverse fields}
\author{Mohammadhossein Dadgar}\email{dadgarmo@msu.edu}
\affiliation{Department of Physics and Astronomy, Michigan State University, East Lansing, Michigan 48824, USA}
\author{Christopher L. Baldwin}\email{baldw292@msu.edu}
\affiliation{Department of Physics and Astronomy, Michigan State University, East Lansing, Michigan 48824, USA}

\date{\today}

\begin{abstract}
Fully-connected transverse interactions have been considered as catalysts for quantum annealing that could mitigate exponentially small gaps and circumvent first-order phase transitions, but their experimental implementation remains challenging.
In this work, we introduce a procedure for emulating their effects via a self-consistent transverse field.
All that is required beyond conventional transverse-field annealing is the ability to make measurements in the transverse ($\hat{\sigma}^x$) basis.
We show that this protocol yields identical dynamics in the large-system limit, and study the approach to that limit in numerical simulations of the (uniform) $p$-spin model.
However, realizing the protocol in practice requires us to consider a series of approximate variants, each of whose errors we quantify and demonstrate can be made sufficiently small.
Lastly, we show how to map the protocol onto annealing platforms that vary only a single control parameter.
Even after the multiple stages of approximation, our procedure can generate dynamics that agree well with the original transverse-interaction catalyst, establishing self-consistent transverse fields as a viable alternative on near-term quantum annealers.
\end{abstract}

\maketitle

\section{Introduction} \label{sec:intro}

Quantum annealing (QA) --- the use of quantum fluctuations to locate the ground states of difficult optimization problems and spin glass models~\cite{AdiabaticQuantumComputation2018albash,Hauke2020Perspectives,Rajak2023Quantum} --- has been generalized repeatedly since its original formulation using a transverse field~\cite{QuantumAnnealingTransverse1998kadowaki,Farhi2001Quantum,Santoro2002Theory}.
Prominent variants of QA include inhomogeneous annealing~\cite{ExponentialSpeedupQuantum2018susa,Adame2020Inhomogeneous,AnomalouslySlowDynamics2025dadgar}, reverse annealing~\cite{ReverseAnnealingFully2018ohkuwa,Marshall2019Power,Mehta2025Unraveling,Baldwin2025Simulated,Le2025Adiabatic}, incorporation of counterdiabatic driving~\cite{Passarelli2020Counterdiabatic,Passarelli2023Counterdiabatic}, and the use of transverse interactions as ``catalysts''.
This latter variant has been particularly encouraging, since early theoretical works proved that the inclusion of fully-connected transverse interactions can convert first-order phase transitions during the annealing protocol (which typically require exponential time to pass through) into second-order transitions (which typically require only polynomial time) in solvable toy models~\cite{QuantumAnnealingAntiferromagnetic2012seki,ManybodyTransverseInteractions2012seoane,QuantumAnnealingAntiferromagnetic2015seki}.
A number of studies have since explored the potential advantages of transverse-interaction catalysts both numerically and analytically~\cite{NonstoquasticHamiltoniansQuantum2017hormozi,RelationQuantumFluctuations2017susa,Albash2019Role,Takada2020MeanField,Takada2021Phase,EffectsXXCatalysts2024feinstein,Nutricati2025Enhancing,Ghosh2026Enhancement}.

However, an issue with transverse interactions is that they are quite difficult (although not impossible~\cite{DemonstrationNonstoquasticHamiltonian2020ozfidan}) to realize experimentally in QA platforms.
In this work, we show how the effects of fully-connected transverse interactions can instead be emulated by adjusting the transverse field self-consistently, thus providing a path towards implementing the catalysts on present-day devices.
In fact, the only capability that our self-consistent protocol requires beyond conventional transverse-field QA --- measurements in the transverse basis --- has recently been demonstrated on the D-Wave quantum annealers~\cite{Deshpande2026AnalogDigital}.

\begin{figure}
\centering
\includegraphics[width=0.6\linewidth]{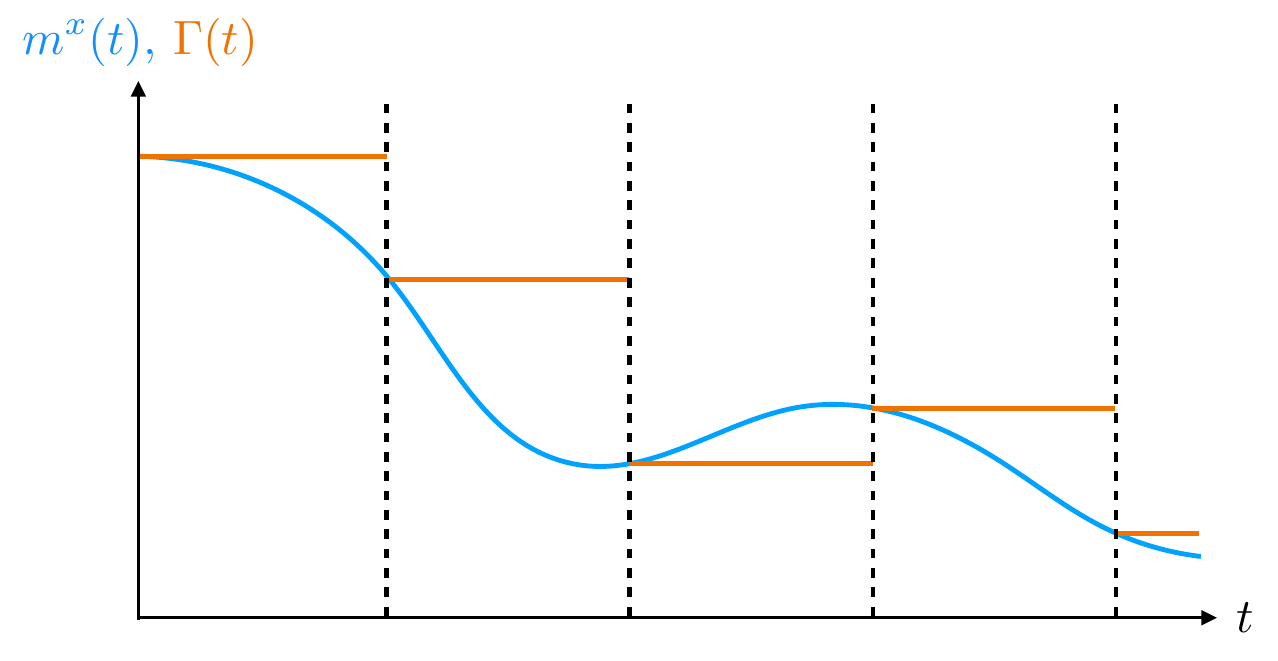}
\caption{Sketch of the self-consistent protocol for emulating fully-connected transverse interactions. We add an additional transverse field $\Gamma(t)$ to the Hamiltonian (orange curve) instead of the transverse interactions. At equally spaced intervals (dashed lines), we estimate the average $x$-magnetization (blue curve) and set $\Gamma(t)$ equal to that value for the subsequent interval. Note that more sophisticated interpolation schemes for $\Gamma(t)$ are also possible (see Appendix~\ref{sec:alternate_interpolation}).}
\label{fig:protocol_cartoon}
\end{figure}

The rationale for this self-consistent protocol comes from mean-field theory, in which one replaces the interactions in a spin Hamiltonian by a field whose strength is governed by the average magnetization.
This ``mean-field decoupling'' has been used many times as a tool for computing phase diagrams in QA studies~\cite{Jorg2010Energy,Bapst2012On,QuantumAnnealingAntiferromagnetic2012seki,ManybodyTransverseInteractions2012seoane,QuantumAnnealingAntiferromagnetic2015seki,QuantumMonteCarlo2017ohzeki,ExponentialSpeedupQuantum2018susa,ReverseAnnealingFully2018ohkuwa}, and it applies to the time evolution as well~\cite{AnomalouslySlowDynamics2025dadgar,Baldwin2025Simulated,Le2025Adiabatic}.
Importantly, the decoupling is exact for fully-connected interactions in the large-$N$ limit (where $N$ denotes the number of qubits in the system).
This is the basis for our self-consistent protocol: by measuring the transverse magnetization after a certain amount of time and using the result to adjust the transverse field at subsequent times, we can reproduce the effects of fully-connected transverse interactions without needing to implement them explicitly.
A cartoon of this procedure is sketched in Fig.~\ref{fig:protocol_cartoon}, and the precise formulation is described in Sec.~\ref{sec:model}.

The bulk of the paper considers the approximations needed to realize this protocol in practice and analyzes their consequences.
In short, the errors introduced by these approximations are all controllable, at least in principle.
One source of error is simply that the mean-field decoupling is not exact for any finite-size system, but the mean-field theory shows that this discrepancy becomes arbitrarily small as the problem size increases.
Further issues are that one can only adapt the transverse field a finite number of times and must base that adjustment on a finite number of measurements, but we show that these errors can be made sufficiently small as well.
Also note that since the protocol uses projective measurements, it must be restarted after every measurement --- this leads to a quadratic overhead in the time required to implement the protocol, but in return, it avoids the need to make non-destructive (``mid-anneal'') measurements in the middle of an annealing schedule.

The question of how to realize non-native transverse interactions has recently been explored by Ref.~\cite{Banks2025Gadgets} as well.
In many ways, our work is complementary to theirs.
Ref.~\cite{Banks2025Gadgets} realizes transverse interactions perturbatively by introducing constraints with a large energy scale, and thus the size of that energy scale controls the accuracy of their implementation.
Here, by contrast, it is the problem size itself that controls the accuracy --- since the ultimate goal of QA is to solve large-scale problems anyway, requiring large sizes does not amount to much of an additional restriction.
However, the downside is that our self-consistent protocol can only emulate fully-connected transverse interactions, whereas that of Ref.~\cite{Banks2025Gadgets} can realize individual two-body transverse terms (although for technical reasons, the terms cannot overlap).
It will be clear from our analysis that one could generalize our approach to realize interactions that are fully-connected only among a subset of spins (such as in Ref.~\cite{Takada2020MeanField}), but then it is the size of that subset that controls the accuracy, and in the limit of trying to approximate individual transverse interactions via self-consistent fields, there is no means of controlling the accuracy whatsoever (although the self-consistent protocol may still be of interest as a standalone catalyst).

In the following section, we define our self-consistent protocol and its various approximations more precisely.
Then in Sec.~\ref{sec:errors}, we analyze the errors introduced by those approximations.
In Sec.~\ref{sec:single_parameter}, we discuss how the self-consistent protocol can be formulated in terms of a single time-dependent control parameter (in contrast to the original three), and then we conclude in Sec.~\ref{sec:conclusion}.
Further details are given in appendices: Appendix~\ref{sec:appendixb} shows how the self-consistent protocol can be derived from mean-field theory, Appendix~\ref{sec:alternate_interpolation} gives an example of a more sophisticated interpolation scheme for one of the approximation stages, and Appendix~\ref{sec:phase_transitions} demonstrates that the self-consistent protocol continues to work when crossing phase transitions.

\section{Model and protocols} \label{sec:model}

Conventional QA uses the Hamiltonian
\begin{equation} \label{eq:conventional_QA_hamiltonian}
H = s H_0(\hat{\sigma}^z) - (1 - s) \sum_{i=1}^N \hat{\sigma}_i^x,
\end{equation}
where $H_0$ is the problem Hamiltonian whose ground state is desired, usually diagonal in the computational ($\hat{\sigma}^z$) basis ($N$ denotes the number of qubits in the problem).
Starting with the qubits aligned along the transverse field, the parameter $s$ varies in time and increases from 0 to 1 during the protocol --- it is simplest to take $s(t) = t/T$, where $T$ denotes the protocol runtime.

QA with a transverse-interaction catalyst adds a third term to the Hamiltonian, consisting of $XX$ interactions between all pairs of spins (normalized by $N$ so as not to dominate the other terms):
\begin{equation} \label{eq:ED_hamiltonian}
\begin{aligned}
H &= s \lambda H_0(\hat{\sigma}^z) +  \frac{s (1 - \lambda)}{N} \sum_{i,j=1}^N \hat{\sigma}_i^x \hat{\sigma}_j^x - (1 - s) \sum_{i=1}^N \hat{\sigma}_i^x.
\end{aligned}
\end{equation}
The additional parameter $\lambda$ also varies in time and increases from 0 to 1, but now we have more freedom in the path that the protocol traces through the $(s, \lambda)$ plane.
For simplicity, we will usually consider $s(t) = \lambda(t) = t/T$.

The focus of this paper is not on whether and how Eq.~\eqref{eq:ED_hamiltonian} provides an advantage over Eq.~\eqref{eq:conventional_QA_hamiltonian} for QA, as this has already been investigated in a number of works.
The focus is instead on how to realize Eq.~\eqref{eq:ED_hamiltonian}, and some straightforward generalizations that we discuss in Sec.~\ref{sec:conclusion}, in a platform that only has access to a transverse field and does not have native $XX$ interactions.
We show that this is indeed possible simply by modifying the time dependence of the transverse field self-consistently.
In other words, the dynamics generated by Eq.~\eqref{eq:ED_hamiltonian}, which we refer to as the ``exact dynamics'' (ED) in what follows, can also be generated by the Hamiltonian
\begin{equation} \label{eq:SC_hamiltonian}
H = s\lambda H_0(\hat{\sigma}^z) + \big[  2s (1 - \lambda) \Gamma(t) - (1 - s) \big] \sum_{i=1}^N \hat{\sigma}_i^x,
\end{equation}
for an appropriate choice of $\Gamma(t)$.
In order to be realizable in practice, however, we must consider a hierarchy of approximations to $\Gamma(t)$, each of which gives rise to a corresponding protocol (although only the final protocol, SCM, is experimentally feasible):
\begin{itemize}
\item \textbf{``Self-consistent exact'' protocol (SCE):} $\Gamma(t)$ is the average $x$-magnetization at time $t$, i.e.,
\begin{equation} \label{eq:SCE_definition}
\Gamma_{\textrm{SCE}}(t) = \frac{1}{N} \sum_{i=1}^N \big< \hat{\sigma}_i^x(t) \big>.
\end{equation}
\item \textbf{``Self-consistent discrete'' protocol (SCD):} $\Gamma(t)$ is still the average $x$-magnetization but only updated at multiples of some $w$, i.e., for $t \in [nw, (n+1)w)$,
\begin{equation} \label{eq:SCD_definition}
\Gamma_{\textrm{SCD}}(t) = \frac{1}{N} \sum_{i=1}^N \big< \hat{\sigma}_i^x(nw) \big>.
\end{equation}
\item \textbf{``Self-consistent measurement'' protocol (SCM):} $\Gamma(t)$ is determined by making a finite number $k$ of measurements of the $x$-magnetization (still only updated at multiples of $w$), i.e.,
\begin{equation} \label{eq:SCM_definition}
\Gamma_{\textrm{SCM}}(t) = \frac{1}{Nk} \sum_{i=1}^N \sum_{j=1}^k s_{i(j)}^x(nw),
\end{equation}
where each $s_{i(j)}^x(nw)$ is a measurement of $\hat{\sigma}_i^x$ at time $nw$.
\end{itemize}
We now explain each of these in turn.

SCE is in fact exact for protocols of fixed runtime $T$ in the large-$N$ limit, i.e., Eq.~\eqref{eq:SC_hamiltonian} with Eq.~\eqref{eq:SCE_definition} yields the exact same expectation values for observables as does Eq.~\eqref{eq:ED_hamiltonian}.
We give a formal proof, following the approach used in Refs.~\cite{AnomalouslySlowDynamics2025dadgar,Baldwin2025Simulated,Le2025Adiabatic}, in Appendix~\ref{sec:appendixb}.
The proof is not quite rigorous, however, and so in what follows, we demonstrate that the dynamics under SCE does indeed converge to that of the original Hamiltonian at large $N$.
As a result, one can apply the SCE protocol as an \textit{approximation} to the transverse-interaction catalyst for arbitrary values of $N$ and $T$, and our results establish that there is a limit (large $N$ at fixed $T$) in which it becomes arbitrarily accurate.

Since determining $\langle \hat{\sigma}_i^x(t) \rangle$ requires measuring the $x$-magnetization, and each measurement requires re-initializing the spins and re-evolving to time $t$ (as the wavefunction is collapsed by each measurement), we cannot update $\Gamma(t)$ at every instant of time.
We must introduce a ``waiting time'' $w > 0$ and only update the field at multiples of $w$.
Eq.~\eqref{eq:SCD_definition} gives the simplest way to do so: $\Gamma_{\textrm{SCD}}(t)$ is a step function whose value changes in accordance with the current $x$-magnetization at each multiple of $w$.
More elaborate interpolations that make $\Gamma_{\textrm{SCD}}(t)$ smoother are of course possible, but these do not change the qualitative idea (see Appendix~\ref{sec:alternate_interpolation}).
While SCD becomes arbitrarily accurate in the limit $w \rightarrow 0$, the runtime of the protocol diverges as $w \rightarrow 0$ (see Eq.~\eqref{eq:SCM_runtime} below), and so one must choose a compromise for $w$ in practice.

Lastly, we of course cannot determine $\langle \hat{\sigma}_i^x(t) \rangle$ exactly, but rather must make a finite number $k$ of measurements of $\hat{\sigma}_i^x$ and take the sample mean of those outcomes.
This gives us the SCM protocol, which can in principle be carried out on experimental platforms (although it may be challenging to implement the required time-dependence of the control field).
Note that SCM does still require one capability beyond what is needed for conventional QA: measurement in non-computational (but still product-state) bases.
Coincidentally, this has recently been shown to be possible on current D-Wave hardware~\cite{Deshpande2026AnalogDigital}.

\begin{figure}
\centering
\includegraphics[width=0.5\linewidth]{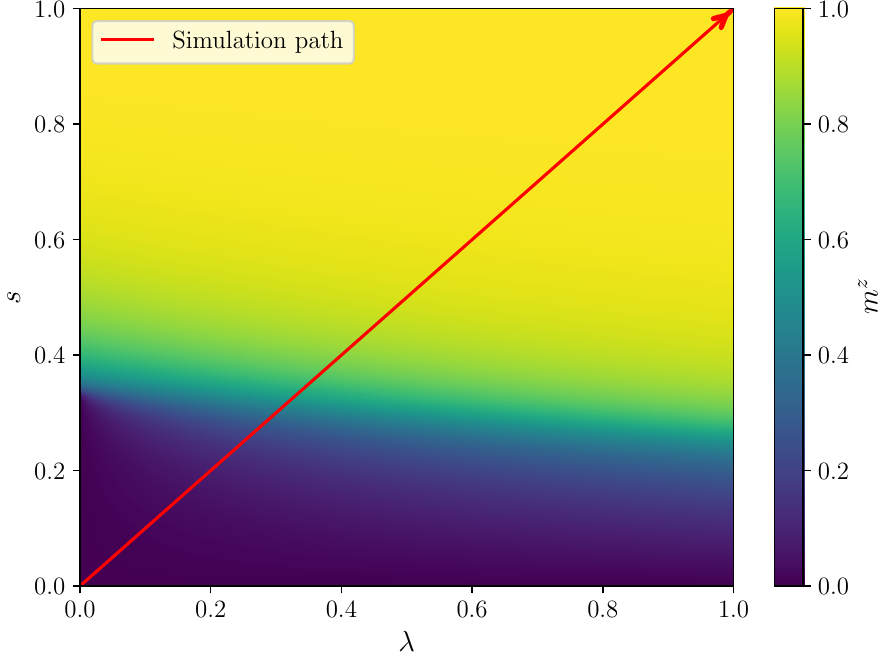}
\caption{Ground-state $z$-magnetization in the $p$-spin model with a transverse field and transverse interactions (i.e., Eq.~\eqref{eq:ED_hamiltonian} using Eq.~\eqref{eq:p_spin_model} for $H_0$), specifically for $p = 3$ and $h = 1$. See Appendix~\ref{sec:phase_transitions} for details of the calculation. The red line indicates the protocol path that we follow in the numerical simulations --- note that no phase transitions occur along that path or anywhere in the parameter space.}
\label{fig:p3}
\end{figure}

As mentioned above, each measurement of the $x$-magnetization collapses the wavefunction, and thus requires re-initializing and re-evolving the spins to time $t$.
We do this $k$ times for each value of $t$ at which we update $\Gamma(t)$, which is every multiple of $w$.
The total runtime of the SCM protocol is therefore not simply the desired evolution time $T$ but rather
\begin{equation} \label{eq:SCM_runtime}
T_{\textrm{tot}} = kw + 2kw + \cdots + k(T - w) \sim \frac{kT^2}{2w}.
\end{equation}
This quadratic overhead should be kept in mind when weighing the benefits and drawbacks of our self-consistent protocol.

The bulk of this paper investigates the errors introduced to the self-consistent approach by finite $N$, $w$, and $k$.
We numerically solve the Schrodinger equation, $i \partial_t | \Psi(t) \rangle = H(t) | \Psi(t) \rangle$, starting from $| \Psi(0) \rangle = | \rightarrow \; \rangle^{\otimes N}$ and using $s(t) = \lambda(t) = t/T$, for both ED and the various self-consistent protocols.
For ED (Eq.~\eqref{eq:ED_hamiltonian}), this amounts to choosing a small time-step $\Delta t$ --- we use $\Delta t = 10^{-3}$ throughout --- and repeatedly acting on $| \Psi(t) \rangle$ with $e^{-i H(t) \Delta t}$.
For each self-consistent protocol, we act on $| \Psi(t) \rangle$ with $e^{-i H(t) \Delta t}$, update the additional field $\Gamma(t)$ using the corresponding formula (Eqs.~\eqref{eq:SCE_definition} through~\eqref{eq:SCM_definition}), and repeat.
Note that we always update $\Gamma(t)$ using the $x$-magnetization \textit{for the same protocol}.

For the problem Hamiltonian $H_0$, we use the $p$-spin model that is common in theoretical studies of QA:
\begin{equation} \label{eq:p_spin_model}
H_0 = -N \left( \frac{1}{N} \sum_{i=1}^N \hat{\sigma}_i^z \right)^p - h \sum_{i=1}^N \hat{\sigma}_i^z.
\end{equation}
Although not a hard optimization problem in of itself, this model has the advantage that we can simulate much larger system sizes $N$: the full Hamiltonian $H(t)$ (for all protocols) can be written solely in terms of the total spin $\hat{S}^{\alpha} \equiv \sum_i \hat{\sigma}_i^{\alpha}$ ($\alpha \in \{x, y, z\}$), and so given our initial state, the dynamics is restricted to the $(N+1)$-dimensional subspace of total spin $N/2$.

Since our focus here is not on the advantages of transverse-interaction catalysts, but rather on the relationship between ED and the self-consistent protocols, we mostly use parameter values that avoid phase transitions for simplicity (hence the inclusion of the longitudinal field in Eq.~\eqref{eq:p_spin_model}).
For the bulk of the paper, we set $p = 3$ and $h = 1$ --- the $z$-magnetization in the ground state of Eq.~\eqref{eq:ED_hamiltonian} evolves smoothly along the path $s(t) = \lambda(t) = t/T$, as shown in Fig.~\ref{fig:p3}.
That said, in Appendix~\ref{sec:phase_transitions}, we consider annealing schedules which do pass through phase transitions (for $p = 5$ and $h = 0$), finding that the qualitative conclusions are the same.

\section{Error between protocols} \label{sec:errors}

In each of the subsections that follow, we compare the dynamics generated by subsequent protocols: ED vs.\ SCE, SCE vs.\ SCD, and SCD vs.\ SCM.
For the simple problem Hamiltonians that we consider here (Eq.~\eqref{eq:p_spin_model}), it is natural to focus on the trajectory of the average $z$-magnetization $m^z(t) \equiv N^{-1} \sum_i \langle \hat{\sigma}_i^z(t) \rangle$, as the expectation value of $H_0$ is uniquely determined by $m^z(t)$.
To quantify the error between protocols A and B, we use
\begin{equation} \label{eq:error_figure_merit}
\Delta_{\textrm{A/B}}^z \equiv \frac{1}{T} \int_0^T \textrm{d}t \big| m_{\textrm{A}}^z(t) - m_{\textrm{B}}^z(t) \big|.
\end{equation}
Note that since we are interested in how well each subsequent protocol can reproduce the dynamics of the prior one, averaging the difference in magnetization over time is a more natural figure of merit than the final difference $|m_{\textrm{A}}^z(T) - m_{\textrm{B}}^z(T)|$: two trajectories might happen to end at similar values of $m^z(T)$ despite differing appreciably at earlier times.

\subsection{ED vs.\ SCE} \label{subsec:ED_vs_SCE}

\begin{figure}
\centering
\includegraphics[width=\textwidth]{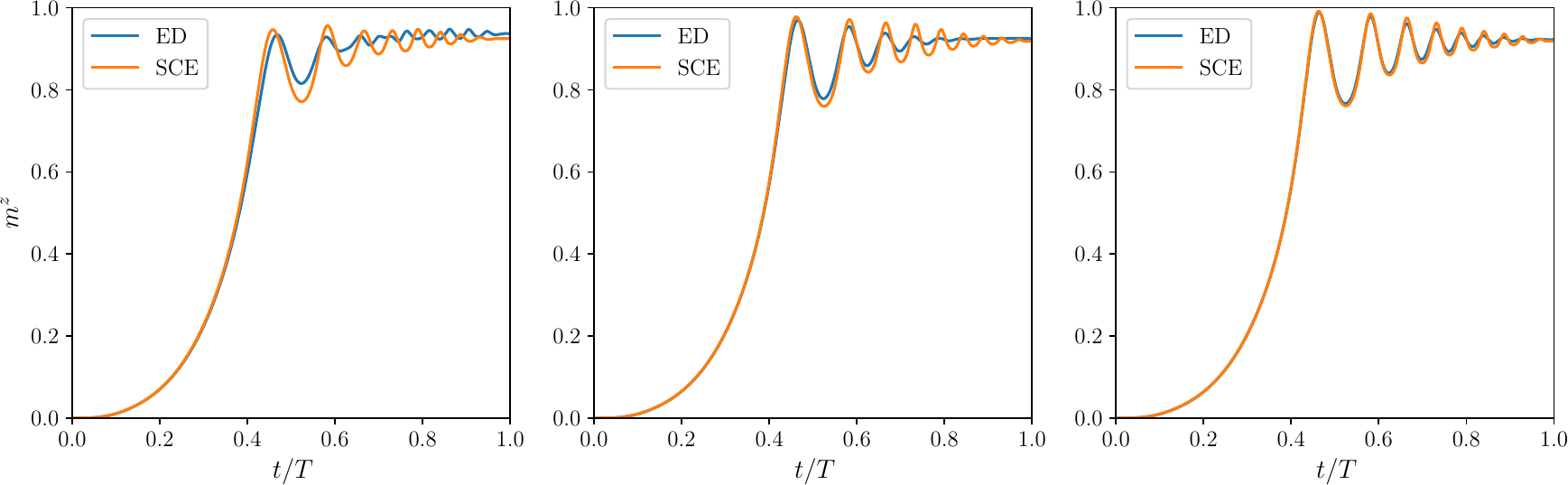}
\caption{Dynamics of the $z$-magnetization under ED (blue) and SCE (orange), for the $p$-spin model with $p = 3$ and $h = 1$. The annealing schedule is $s(t) = \lambda(t) = t/T$, and the runtime is $T = 25$. The three panels differ only in the system size --- $N = 25$ (left), $N = 100$ (center), and $N = 400$ (right).}
\label{fig:SCE_examples}
\end{figure}

Recall that SCE replaces the transverse interactions by a transverse field whose strength is proportional to the average $x$-magnetization at that moment.
The two are equivalent in the large-$N$ limit (holding $T$ fixed), but discrepancies appear at finite $N$.
Fig.~\ref{fig:SCE_examples} shows three examples comparing the $z$-magnetization under ED and SCE, for the same $H_0$ ($p = 3$ and $h = 1$) and the same value of $T$, while increasing $N$.
The two trajectories are qualitatively similar but do not agree in any specific features at the smallest size ($N = 25$).
They agree much better at $N = 100$, although the oscillations in $m^z(t)$ are noticeably longer-lived under SCE.
Lastly, the trajectories clearly agree extremely well by $N = 400$, not only qualitatively but quantitatively (the error is approximately $4 \times 10^{-3}$).

\begin{figure}
\centering
\includegraphics[width=0.48\linewidth]{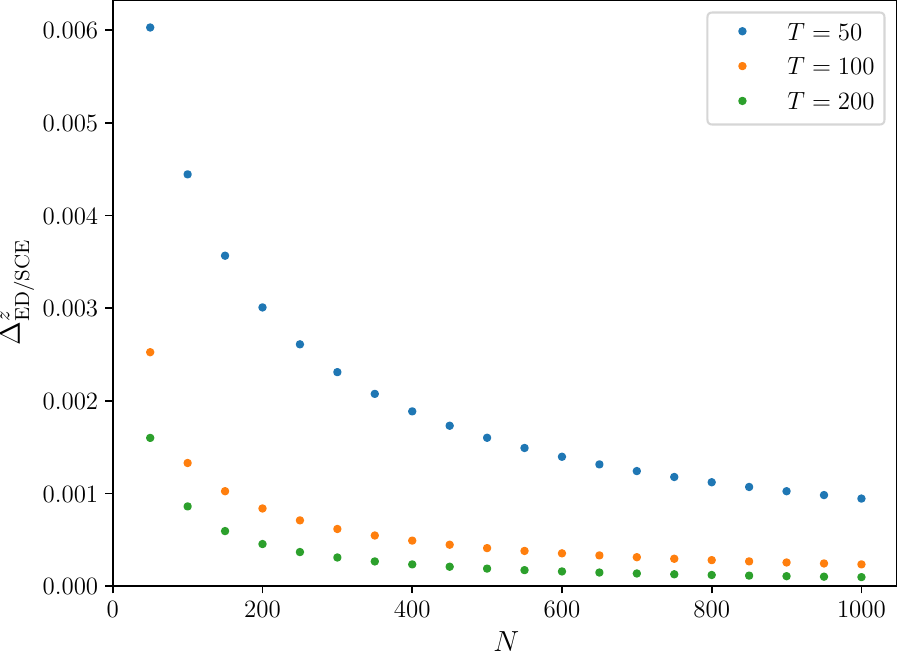}
\quad
\includegraphics[width=0.48\linewidth]{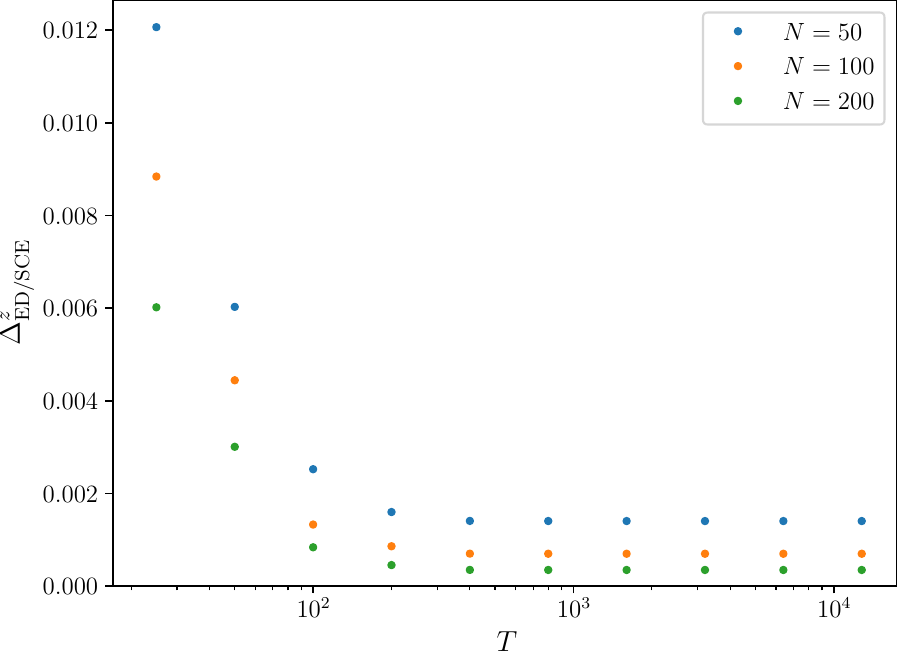}
\caption{Error between ED and SCE, $\Delta_{\textrm{ED/SCE}}^z$, as a function of $N$ (left) and $T$ (right). The problem Hamiltonian again has $p = 3$ and $h = 1$, and the annealing schedule is $s(t) = \lambda(t) = t/T$.}
\label{fig:ED_vs_SCE_errors}
\end{figure}

To be more quantitative, we plot the error $\Delta_{\textrm{ED/SCE}}^z$ versus $N$ in the left panel of Fig.~\ref{fig:ED_vs_SCE_errors}.
As expected, the error decreases monotonically with $N$.
Plotted on a log-log scale (not shown), we find that it scales roughly as $\Delta_{\textrm{ED/SCE}}^z \sim 1/N$.
It is possible that one could derive this scaling from corrections to the saddle-point approximation in Appendix~\ref{sec:appendixb}, but we leave this for future work.

The dependence of $\Delta_{\textrm{ED/SCE}}^z$ on $T$ (right panel of Fig.~\ref{fig:ED_vs_SCE_errors}) is more interesting.
The derivation in Appendix~\ref{sec:appendixb} that relates SCE to ED only holds for $T = O(1)$ with respect to $N$, and so one might expect that the error should increase with $T$.
However, increasing $T$ also makes the dynamics closer to adiabatic and thus reduces the oscillations seen in Fig.~\ref{fig:SCE_examples}.
Since those oscillations are the main source of discrepancy between ED and SCE, one might conversely expect the error to be smaller at larger $T$.
For this specific model, we find that the net effect is for $\Delta_{\textrm{ED/SCE}}^z$ to decrease with $T$ but saturate at a non-zero value.
We caution, however, that this behavior seems to be model-dependent (and even path-dependent within the same model) --- Appendix~\ref{sec:phase_transitions} gives counterexamples where $\Delta_{\textrm{ED/SCE}}^z$ instead increases as a function of $T$.

Regardless of the details, these observations do have one significant implication: strictly speaking, whereas ED can reach the ground state of $H_0$ with vanishing error by taking $T \rightarrow \infty$, SCE cannot --- there is a limit to how closely it can follow the ground state by increasing $T$.
That said, the magnitude of the error can be quite small (note the scale of the $y$-axes in Fig.~\ref{fig:ED_vs_SCE_errors}), and it becomes smaller at larger problem sizes, so the self-consistent protocol could still very well be useful in practice.

\subsection{SCE vs.\ SCD} \label{subsec:SCE_vs_SCD}

\begin{figure}
\centering
\includegraphics[width=0.48\linewidth]{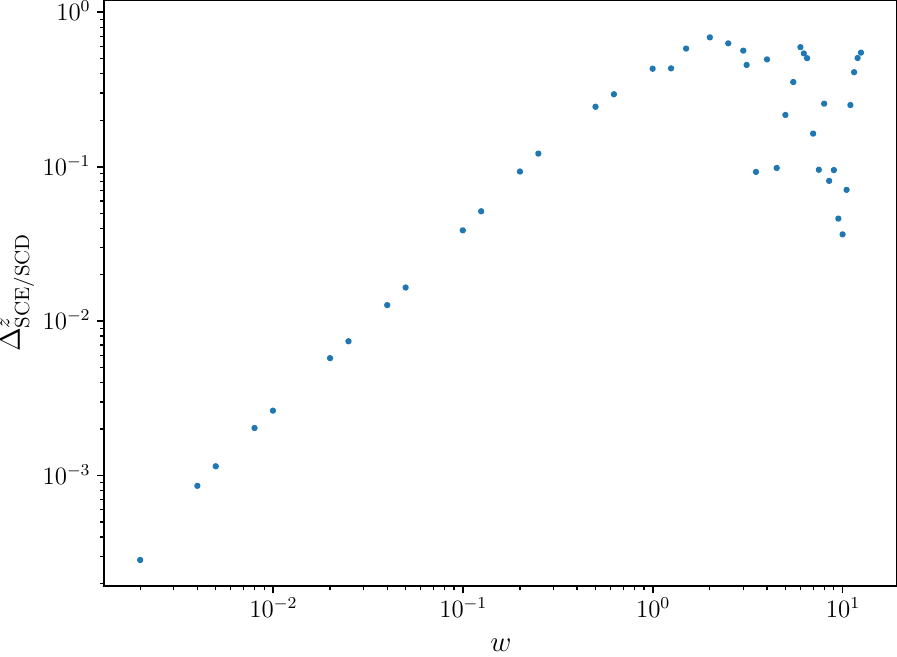}
\quad
\includegraphics[width=0.48\linewidth]{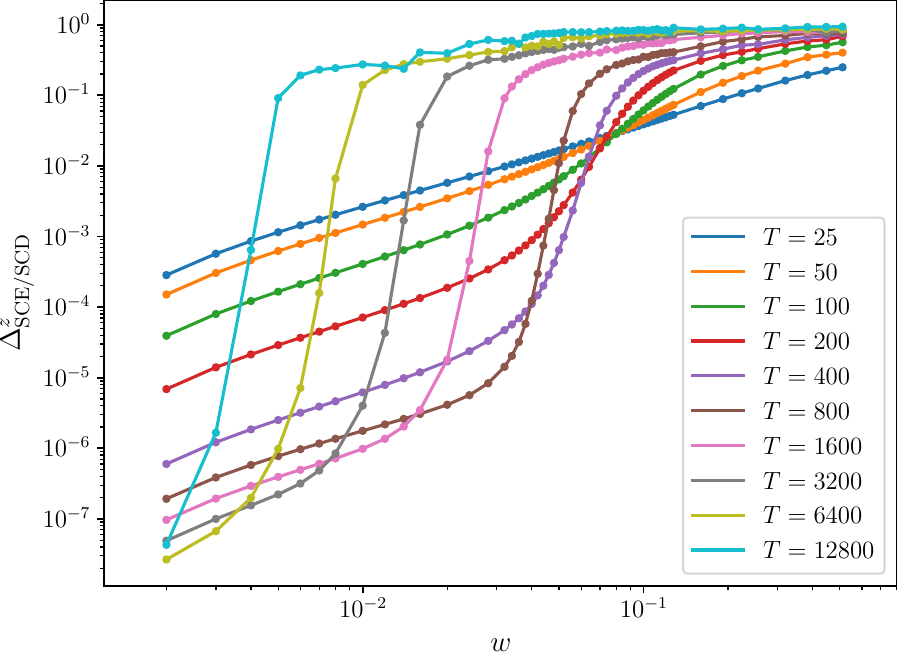}
\caption{Error between SCE and SCD, $\Delta_{\textrm{SCE/SCD}}^z$, as a function of the waiting time $w$ used in SCD (system size is $N = 100$ and the problem Hamiltonian again has $p = 3$ and $h = 1$). The left panel shows $\Delta_{\textrm{SCE/SCD}}^z$ for all values of $w$ considered at $T = 25$, while the right panel is zoomed in to the small-$w$ regime and shows multiple values of $T$.}
\label{fig:SCD0}
\end{figure}

Recall that SCD differs from SCE in that the self-consistent field is updated only at times which are multiples of a new parameter $w$ (SCE results from taking $w \rightarrow 0$).
The left panel of Fig.~\ref{fig:SCD0} shows how the error $\Delta_{\textrm{SCE/SCD}}^z$ depends on this ``waiting time'' $w$.
We see two clear regimes: a small-$w$ regime in which the error scales roughly as $\Delta_{\textrm{SCE/SCD}}^z \sim w$, and a large-$w$ regime in which the error behaves quite erratically.
Note that $\Delta_{\textrm{SCE/SCD}}^z$ by definition cannot be larger than 2, and one might expect the error for uncorrelated protocols to be closer to 1.
Since SCE and SCD start with the same magnetization, the time-averaged error should be slightly reduced further, and so the errors that we find in the large-$w$ regime are close to the maximum expected for uncorrelated protocols.
Nonetheless, it is interesting to observe that there are certain large values of $w$, e.g., $w/T = 0.4$, where the error is anomalously small (although we do not attach much practical significance to these values since the error behaves so unpredictably there).

The distinction between small and large $w$ becomes more substantive by considering the dependence on $T$, as shown in the right panel of Fig.~\ref{fig:SCD0}.
In some aspects, the curves are reminiscent of the finite-size behavior at first-order phase transitions, with $T$ playing the role of system size: the crossover between the small- and large-$w$ regimes becomes sharper as $T$ increases, with the error being extremely small on the small-$w$ side and close to maximal on the large-$w$ side.
However, the crossover region itself drifts towards smaller $w$ as $T$ increases, such that in the large-$T$ limit, any nonzero $w$ will count as ``large'' and SCD will fail to emulate SCE.
This imposes a further restriction, beyond that discussed in Sec.~\ref{subsec:ED_vs_SCE}, on the runtimes of protocols that can be emulated self-consistently.

\begin{figure}
\centering
\includegraphics[width=0.5\linewidth]{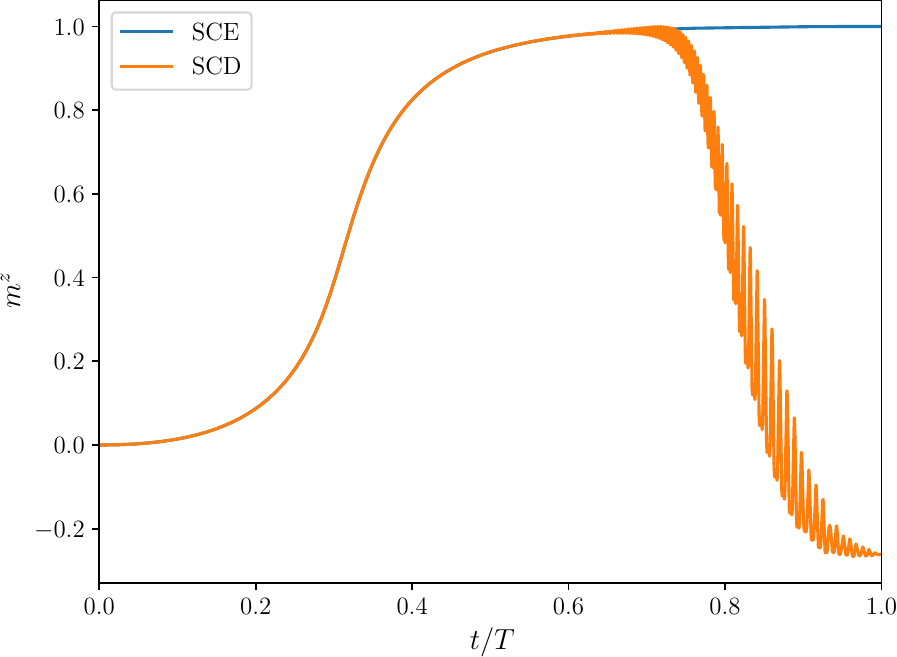}
\caption{Dynamics of the $z$-magnetization under SCE (blue) and SCD (orange). Parameters are $N = 100$, $T = 400$, and $w = 0.1$ (for the same Hamiltonian and annealing schedule as in previous plots).}
\label{fig:esc}
\end{figure}

The reason for these trends becomes clear by comparing the dynamics of the $z$-magnetization under the two protocols, e.g., as shown in Fig.~\ref{fig:esc}.
Relative to SCE, SCD suffers from an instability that causes the trajectory to deviate dramatically but only once the instability has enough time to develop.
Thus to understand whether SCD can emulate SCE, one should compare the protocol runtime to the instability timescale $T_{\textrm{inst}}$: the error will be extremely small if $T < T_{\textrm{inst}}$, but the two protocols will be effectively unrelated if $T > T_{\textrm{inst}}$.

From Fig.~\ref{fig:SCD0}, we see that the location $w_c$ of the crossover scales as a power-law in $T$: $w_c \sim T^{-\alpha}$, with a scaling collapse estimating $\alpha \approx 0.85$.
The inverse then gives how the instability timescale $T_{\textrm{inst}}$ scales with $w$: $T_{\textrm{inst}} \sim w^{-1/\alpha}$, with $1/\alpha \approx 1.18$.
In particular, the timescale diverges as $w \rightarrow 0$.

In Appendix~\ref{sec:alternate_interpolation}, we consider a more sophisticated interpolation scheme for $\Gamma_{\textrm{SCD}}(t)$, such that it changes continuously instead of having discontinuous jumps at every update.
We find the same qualitative behavior as in Fig.~\ref{fig:SCD0}, including the instability that degrades the protocol as $T \rightarrow \infty$, although the timescale for the instability is larger in absolute terms for the same $w$.
This makes sense, as a continuous $\Gamma_{\textrm{SCD}}(t)$ can better approximate the desired $\Gamma_{\textrm{SCE}}(t)$, and it provides an alternate route to addressing the instability rather than simply decreasing $w$.

\subsection{SCD vs.\ SCM} \label{subsec:SCD_vs_SCM}

\begin{figure}
\centering
\includegraphics[width=0.48\linewidth]{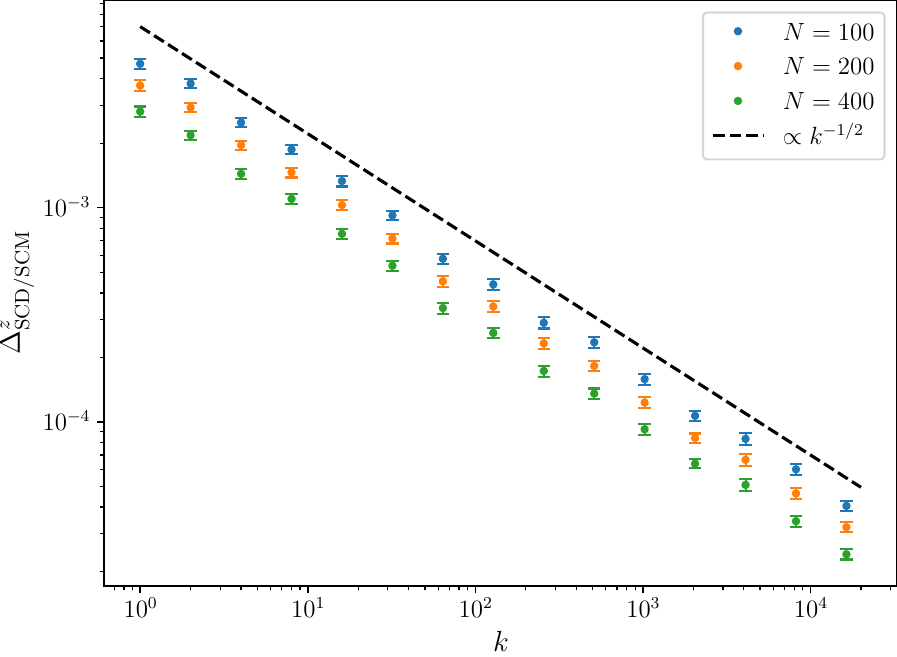}
\quad
\includegraphics[width=0.48\linewidth]{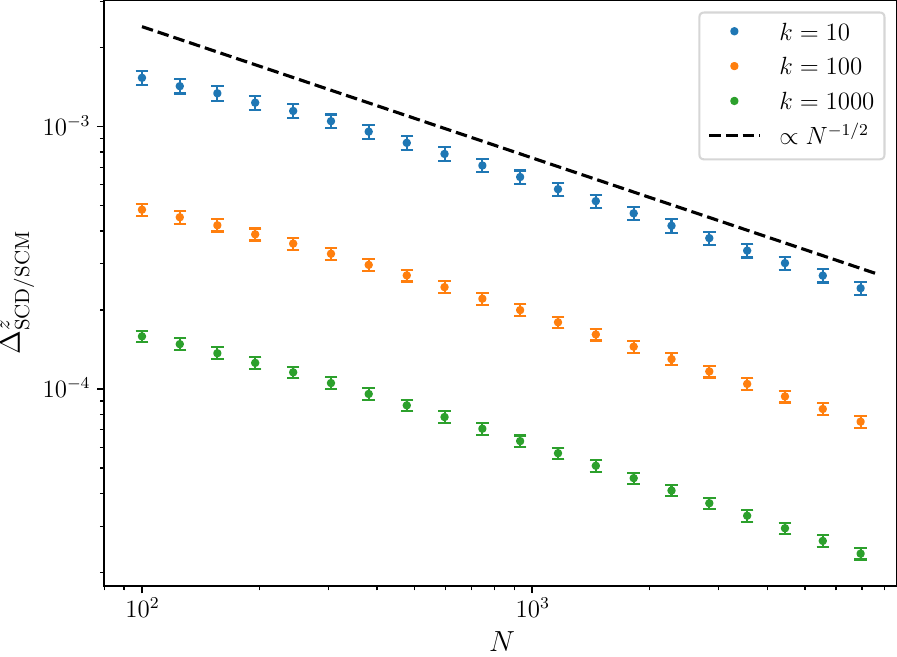}
\caption{Average error between SCD and SCM, $\Delta_{\textrm{SCD/SCM}}^z$, as a function of $k$ (left) and $N$ (right). Other parameters are $T = 25$ and $w = 0.01$ (with the same Hamiltonian and annealing schedule as in previous plots). Note that the $z$-magnetization under SCM, and thus $\Delta_{\textrm{SCD/SCM}}^z$, is a random variable --- the calculation for each point was repeated 100 times (using $k$ measurements each time) to determine the average error and its uncertainty.}
\label{fig:Nk}
\end{figure}

Recall that SCM differs from SCD in that, rather than update the self-consistent field using the expectation value of the $x$-magnetization, we update the field using $k$ measurements of it.
The scaling of the error with both $k$ and the system size $N$ is shown in Fig.~\ref{fig:Nk}, where we see that $\Delta_{\textrm{SCD/SCM}}^z \sim 1/\sqrt{Nk}$.
This behavior is in fact straightforward to understand, as we describe momentarily, but it is nonetheless quite interesting to observe that the error \textit{decreases} with system size --- less measurements are needed to obtain accurate results in larger problems.

It is a classic result in probability that when sampling a random variable $k$ times, the mean of those $k$ measurements differs from the true expectation value by $O(1/\sqrt{k})$, as one can directly confirm by calculating the variance of the sample mean.
In our case, the random variable is the $x$-magnetization $N^{-1} \sum_i \hat{\sigma}_i^x$, whose distribution is given by the wavefunction $| \Psi(t) \rangle$.
Note that this is not quite what is being plotted in Fig.~\ref{fig:Nk}, however: $\Delta_{\textrm{SCD/SCM}}^z$ is the difference in magnetization between two \textit{different} wavefunctions, one evolved under the expectation value of the $x$-magnetization and the other evolved under the sample mean.
Nonetheless, it is quite reasonable to expect that $\Delta_{\textrm{SCD/SCM}}^z$ should inherit the same $O(1/\sqrt{k})$ scaling, and this is indeed what we find in the left panel of Fig.~\ref{fig:Nk}.

Now turn to the $O(1/\sqrt{N})$ scaling.
Recall that for models such as Eq.~\eqref{eq:p_spin_model}, which can be expressed in terms of the total spin $\hat{S}^{\alpha}$, the dynamics is restricted to the subspace of total spin $N/2$.
In this subspace, there is a single basis state $| m \rangle$ having $x$-magnetization $m$, and it is known that the modulus of the wavefunction scales at large $N$ as~\cite{Garg1998Application,Bapst2012On,Baldwin2021Distinct}
\begin{equation} \label{eq:WKB_ansatz}
\big| \langle m | \Psi \rangle \big| \sim \exp{\big[ NS(m) \big]},
\end{equation}
where the function $S(m)$ is independent of $N$.
This result is analogous to the WKB approximation, with $N^{-1}$ playing the role of $\hbar$, and it holds for both the ground state and the time-evolved wavefunction (the exponent $S(m)$ will also depend on $t$ in the latter case).
Since $| \langle m | \Psi \rangle |$ is so sharply peaked at large $N$, the expectation value of the $x$-magnetization will be given simply by the location of the maximum of $S(m)$, denoted $m_0$.
Furthermore, Taylor-expanding $S(m)$ around $m_0$, the zeroth-order term vanishes due to normalization and the first-order term vanishes due to $m_0$ being a maximum, so we are left with
\begin{equation} \label{eq:WKB_ansatz_expanded}
\big| \langle m | \Psi \rangle \big| \sim \exp{\bigg[ -\frac{N |S''(m_0)|}{2} \big( m - m_0 \big)^2 \bigg]}.
\end{equation}
In other words, the wavefunction is approximately a Gaussian with variance of order $1/N$.
Thus a measurement of the $x$-magnetization will differ from the expectation value by $O(1/\sqrt{N})$.

While this argument is specific to fully-connected models, we expect the same scaling in short-range models, assuming only that the wavefunction in question has a finite correlation length $\xi$.
Then measurements of the local magnetization in regions separated by more than $\xi$ will be effectively uncorrelated, meaning that a measurement of the global magnetization amounts to $O(N/\xi^d)$ independent measurements of the local magnetization (where $d$ is the dimensionality).
By the same probability argument used above, the fluctuations in the global magnetization around its expectation value should therefore be $O(1/\sqrt{N})$.

Once again, strictly speaking, these arguments relate the expectation value of the $x$-magnetization to a measurement of it using the same wavefunction, whereas $\Delta_{\textrm{SCD/SCM}}^z$ compares the magnetization under two different wavefunctions.
Yet we expect the scaling of $\Delta_{\textrm{SCD/SCM}}^z$ to be the same, and this is indeed what Fig.~\ref{fig:Nk} shows (at sufficiently large $N$).

These observations explain the scaling $\Delta_{\textrm{SCD/SCM}}^z \sim 1/\sqrt{Nk}$.
The scaling of the error with $T$ and $w$ (not shown) is more complicated.
We have not identified a simple functional form, but the overall trend is that the error increases with both $T$ and $w$.
This makes sense: a longer runtime allows more time for the two protocols to diverge, and similarly, waiting longer between updates of the transverse field allows more time for initial errors to magnify.

\section{Mapping to single-parameter schedules} \label{sec:single_parameter}

There is one final aspect to address: many experimental platforms for quantum annealing, such as the D-Wave machines, do not have separate control over the coefficients of the problem Hamiltonian and the transverse field.
Instead they have a single control parameter $u \in [0, 1]$, and evolve under the Hamiltonian $A(u) H_0 - B(u) \sum_i \hat{\sigma}_i^x$, where $A(u)$ and $B(u)$ are known functions (monotonically increasing and decreasing respectively).
Since our self-consistent protocol adds a contribution $\Gamma$ to the transverse field without subtracting a corresponding amount from $H_0$, a non-trivial relationship between $u$ and $(s, \lambda, \Gamma)$ is required in order to implement the protocol on such platforms.

To be concrete, suppose we have already determined $\Gamma(t)$ up to time $nw$.
We want to evolve under the ``protocol'' Hamiltonian in Eq.~\eqref{eq:SC_hamiltonian}, namely
\begin{equation} \label{eq:single_parameter_desired_Hamiltonian}
\begin{aligned}
H_{\textrm{prot}}(t) &= s(t) \lambda(t) H_0 + \Big[ 2 s(t) \big( 1 - \lambda(t) \big) \Gamma(t) - 1 + s(t) \Big] \sum_i \hat{\sigma}_i^x \\
&= s(t) \lambda(t) \Bigg( H_0 - \frac{1 - s(t) - 2 s(t) (1 - \lambda(t)) \Gamma(t)}{s(t) \lambda(t)} \sum_i \hat{\sigma}_i^x \Bigg),
\end{aligned}
\end{equation}
so that we can measure the $x$-magnetization at time $nw$ and thus determine $\Gamma(t)$ up to time $(n+1) w$.
However, we are forced to do so using the ``physical'' Hamiltonian
\begin{equation} \label{eq:single_parameter_actual_Hamiltonian}
\begin{aligned}
H_{\textrm{phys}}(t) &= A \big( u(t) \big) H_0 - B \big( u(t) \big) \sum_i \hat{\sigma}_i^x \\
&= A \big( u(t) \big) \Bigg( H_0 - \frac{B(u(t))}{A(u(t))} \sum_i \hat{\sigma}_i^x \Bigg).
\end{aligned}
\end{equation}
We must determine the appropriate schedule $u(t)$ given $s(t)$, $\lambda(t)$, and $\Gamma(t)$ as input.

Comparing the final expressions in Eqs.~\eqref{eq:single_parameter_desired_Hamiltonian} and~\eqref{eq:single_parameter_actual_Hamiltonian}, it is clear that to match the strength of the transverse field at a given point in the protocol, we should set $u$ equal to the value for which $B(u)/A(u) = [1 - s - 2s (1 - \lambda) \Gamma]/s \lambda$.
Note that this equation always has a unique solution, which can easily be determined numerically, as long as $A(u)$ is monotonically increasing from 0 and $B(u)$ is monotonically decreasing (although the required $B(u)$ could conceivably be negative, which may not be possible on certain platforms).
However, we should not determine $u(t)$ using the values of $s$, $\lambda$, and $\Gamma$ at the same time $t$, since the overall coefficients of the Hamiltonians in Eqs.~\eqref{eq:single_parameter_desired_Hamiltonian} and~\eqref{eq:single_parameter_actual_Hamiltonian} are different.
Let $\tau(t)$ denote the time at which we evaluate $s$, $\lambda$, and $\Gamma$ to determine $u$ at time $t$ --- in order for the physical Hamiltonian to generate the same dynamics as the protocol Hamiltonian, we must have $H_{\textrm{phys}}(t) \textrm{d}t = H_{\textrm{prot}}(\tau) \textrm{d}\tau$ at all times.
Comparing the coefficients of $H_{\textrm{phys}}$ and $H_{\textrm{prot}}$, this gives a differential equation relating $t$ to $\tau$:
\begin{equation} \label{eq:single_parameter_time_stretching}
\frac{\textrm{d}\tau}{\textrm{d}t} = \frac{A(u(t))}{s(\tau) \lambda(\tau)}.
\end{equation}

\begin{figure}
\centering
\includegraphics[width=0.48\linewidth]{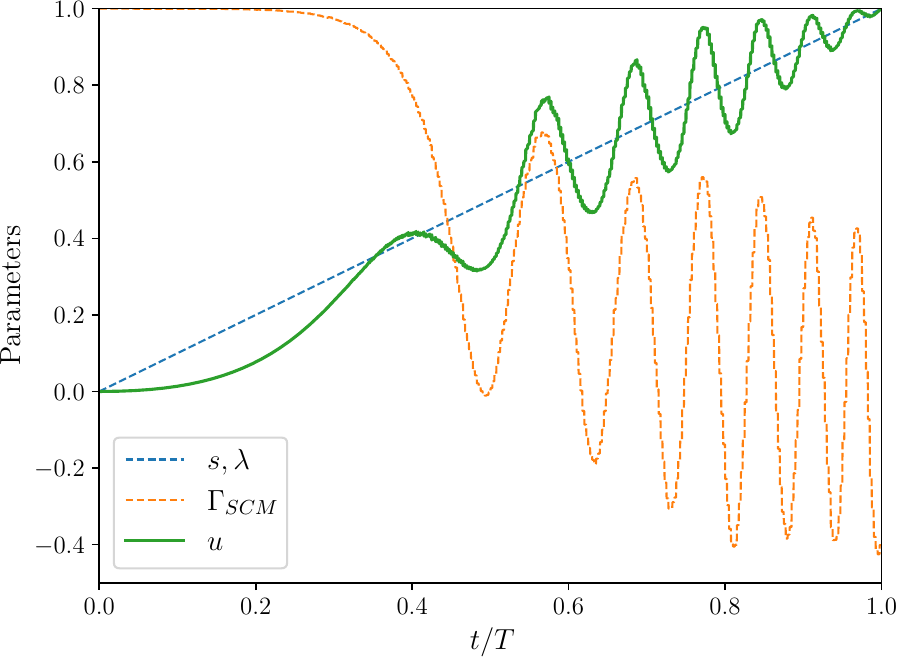}
\caption{Example of the single-parameter schedule $u(t)$ that results from a specified $s(t)$ and $\lambda(t)$ (using the same Hamiltonian as previously with parameters $N = 100$, $T = 20$, $w = 0.05$, $k = 1000$). The dashed blue line shows $s(t)$ and $\lambda(t)$, both chosen to be simply $t/T$. Dashed orange shows the resulting self-consistent field under SCM, and solid green shows $u(t)$. For simplicity, the calculations were performed using $A(u) = u$ and $B(u) = 1-u$.}
\label{fig:single_parameter_schedule}
\end{figure}

To summarize, we determine the schedule $u(t)$ by solving
\begin{equation} \label{eq:single_parameter_effective_value}
\frac{B(u(t))}{A(u(t))} = \frac{1 - s(\tau) - 2s(\tau) (1 - \lambda(\tau)) \Gamma(\tau)}{s(\tau) \lambda(\tau)},
\end{equation}
where $\tau(t)$ is obtained by integrating Eq.~\eqref{eq:single_parameter_time_stretching}.
An example of the schedule that results from this calculation is shown in Fig.~\ref{fig:single_parameter_schedule} (where for simplicity, we use $A(u) = u$ and $B(u) = 1 - u$).

\section{Conclusion} \label{sec:conclusion}

In this work, we have introduced a self-consistent protocol, abbreviated the SCM protocol, to emulate fully-connected transverse interactions in quantum annealing using only a transverse field.
The protocol requires multiple stages of approximation: first replacing the transverse interactions by a field equal to the average $x$-magnetization at that moment, then updating the field not at every instant but a finite number of times, and lastly estimating the magnetization using only a finite number of measurements.
Via numerical simulations on the $p$-spin model, we have shown that the error at each stage can be made suitably small by controlling a corresponding parameter --- system size $N$ at the first, waiting time $w$ at the second, and number of measurements $k$ at the third --- thus demonstrating the feasibility of our protocol on experimental platforms.
In addition, we have discussed how to implement the protocol on devices which only have a single control parameter tuning the relative strength of the problem Hamiltonian and transverse field.

\begin{figure}
\centering
\includegraphics[width=0.48\linewidth]{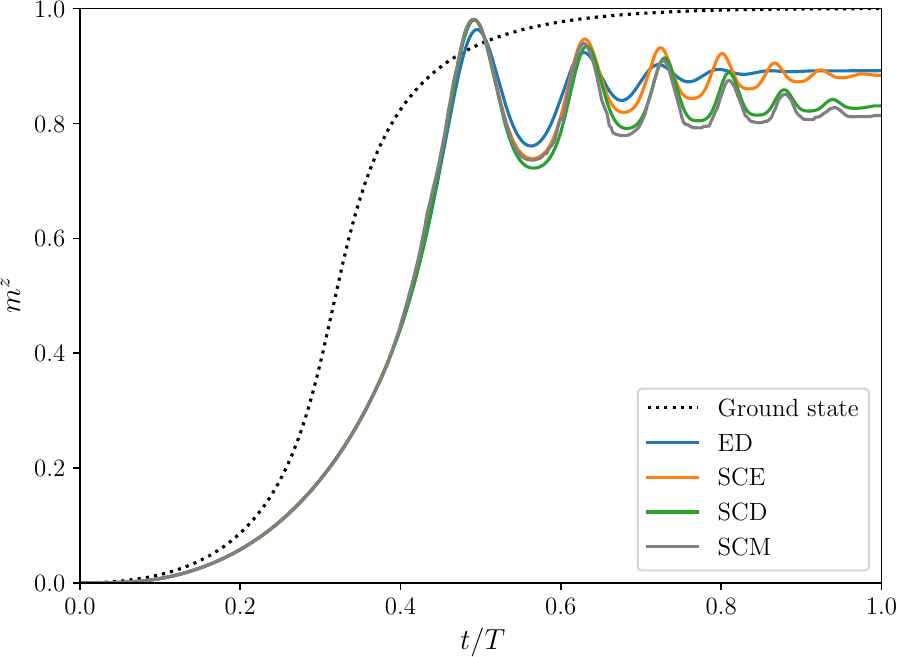}
\caption{Dynamics of the $z$-magnetization under ED and the various self-consistent protocols. Parameters are $N = 100$, $T = 20$, $w = 0.05$, $k=1$ (for the same Hamiltonian and annealing schedule as in previous plots). The $z$-magnetization of the instantaneous ground state during the protocol is shown for reference (dashed line).}
\label{fig:comparison}
\end{figure}

Since we have considered the error introduced at each stage of approximation separately, one might worry that their cumulative effect makes the protocol impractical.
Thus in Fig.~\ref{fig:comparison}, we give an example comparing the $z$-magnetization under the transverse interactions (blue) with all stages of the self-consistent protocol (orange, green, and ultimately gray).
We specifically use parameter values that give good but not perfect agreement, to demonstrate how far one can push the approximations.
Note that, using Eq.~\eqref{eq:SCM_runtime}, the overhead in total runtime is $T_{\textrm{tot}}/T = 200$.
This is not particularly burdensome, since the overhead is not analogous to performing a single anneal over a much longer timescale (where one would have to worry about the coherence time of the system), but rather analogous to running 200 sweeps on the original timescale.

There are two particularly interesting features of the self-consistent protocol that are on full display in Fig.~\ref{fig:comparison}.
First, having a non-zero waiting time $w$ between updates introduces an instability into the dynamics --- in order for the instability not to manifest during the protocol, the number of updates $T/w$ needs to be rather large (Fig.~\ref{fig:comparison} has $T/w = 400$).
Second, as explained in Sec.~\ref{subsec:SCD_vs_SCM}, very few measurements are required at each update in sufficiently large problems --- note that in Fig.~\ref{fig:comparison}, we are even able to obtain reasonable agreement making only a \textit{single} measurement per update.

In general, one important limitation of the self-consistent protocol is that it cannot return the exact ground state with probability 1 by taking $T \rightarrow \infty$, in contrast to conventional annealing (with or without the transverse interactions).
There are two issues that arise in this limit.
First, the original mapping from transverse interactions to a self-consistent field breaks down at large $T$ --- more precisely, it only holds for $T = O(1)$ with respect to $N$.
We have found that in practice, this imposes a limit on how low in energy the self-consistent protocol can reach while increasing $T$.
Second, we must take care to avoid the aforementioned instability caused by non-zero $w$.
The timescale for this instability diverges as $w \rightarrow 0$, and so this issue can be mitigated for any $T$ by making $w$ sufficiently small (or by improving the interpolation scheme used between measurements).
Yet on the other hand, when taking $T \rightarrow \infty$ while holding $w$ fixed, the instability will ultimately kick in and degrade the performance of the protocol.

While this work has explicitly studied antiferromagnetic two-body transverse interactions, there are a number of straightforward generalizations to which our self-consistent protocol applies equally well.
The simplest case is emulating ferromagnetic transverse interactions, which one can do simply by changing the sign of the self-consistent field $\Gamma(t)$.
One can also emulate higher-body (still fully-connected) transverse interactions by modifying the relationship between $\Gamma(t)$ and the $x$-magnetization.
For example, to realize $q$-body interactions, the SCE protocol would have $\Gamma_{\textrm{SCE}}(t) \propto (N^{-1} \sum_i \langle \sigma_i^x(t) \rangle)^{q-1}$ (with the corresponding expressions for $\Gamma_{\textrm{SCD}}(t)$ and $\Gamma_{\textrm{SCM}}(t)$).
Lastly, note that the self-consistent protocol also applies to transverse interactions that are not fully-connected among all $N$ spins but merely connected among subsets of $\Lambda$ spins --- in that case, it is the size of $\Lambda$ that controls the accuracy of the protocol, and one would need a different self-consistent contribution for each connected subset of spins.
We have done preliminary calculations suggesting that the same conclusions apply to these various generalizations, but we leave a systematic study for future work.

Taking this last observation further, one could in principle use self-consistent fields as an approximation to \textit{arbitrary} transverse interactions: for each qubit, replace the $XX$ interactions involving that qubit with a transverse field proportional to the $x$-magnetization of the neighboring spins.
One difficulty with this approach is that one would need separate control over the field acting on each individual qubit.
More significantly, the approximation becomes uncontrolled, i.e., there is no parameter that one could vary to make the error between the self-consistent protocol and the original Hamiltonian arbitrarily small.
In this more general context, it is better to view self-consistent fields as a standalone type of catalyst, one that is motivated by but independent of transverse interactions.
It would be quite informative to investigate the benefits that such self-consistent transverse fields may provide.

\section{Acknowledgements} \label{sec:acknowledgements}

This work was supported by the U.S. National Science Foundation under award No.~2508604.
Simulations were performed using the high-performance computing center of the Institute for Cyber-Enabled Research (ICER) at Michigan State University.

\bibliography{ref}

\appendix

\section{Derivation of the self-consistent exact protocol}\label{sec:appendixb}

Here we give a formal proof that the SCE protocol produces exactly the same dynamics as the original Hamiltonian in the large-$N$ limit (while holding all other parameters, including the runtime $T$, constant).
To reiterate, the original Hamiltonian is
\begin{equation} \label{eq:proof_original_Hamiltonian}
H(t) = s(t) \lambda(t) H_0 + \frac{s(t) (1 - \lambda(t))}{N} \sum_{i,j=1}^N \hat{\sigma}_i^x \hat{\sigma}_j^x - (1 - s(t)) \sum_{i=1}^N \hat{\sigma}_i^x.
\end{equation}
Note that the following proof applies to a completely arbitrary problem Hamiltonian $H_0$ --- it is only the fully-connected nature of the transverse interactions that we require.

We consider the generating functional for the dynamics of the system, as discussed in references on non-equilibrium many-body theory~\cite{Rammer2007,Kamenev2011,Sieberer2016Keldysh}:
\begin{equation} \label{eq:proof_generating_functional}
\mathcal{Z}[\xi] \equiv \textrm{Tr} \Big( \mathcal{T} e^{-i \int_0^T \textrm{d}t [H(t) - \sum_i \vec{\xi}_i^+(t) \cdot \vec{\sigma}_i]} \Big) \rho(0) \Big( \mathcal{T} e^{-i \int_0^T \textrm{d}t [H(t) - \sum_i \vec{\xi}_i^-(t) \cdot \vec{\sigma}_i]} \Big)^{\dag},
\end{equation}
where $\rho(0)$ is the initial density matrix and $\mathcal{T}$ denotes time-ordering.
Note that when $\xi = 0$, the generating functional is simply the trace of the final density matrix, i.e., equal to one.
Thus its numerical value is not of interest.
Instead, it is useful because derivatives with respect to $\xi$ yield expectation values of observables at the corresponding times: for example,
\begin{equation} \label{eq:proof_generating_functional_derivatives}
-i \frac{\delta \mathcal{Z}[\xi]}{\delta \xi_i^{+z}(t)} \Bigg|_{\xi = 0} = \big< \hat{\sigma}_i^z(t) \big>, \qquad \frac{\delta^2 \mathcal{Z}[\xi]}{\delta \xi_i^{-x}(t) \delta \xi_j^{+y}(t')} \Bigg|_{\xi = 0} = \big< \hat{\sigma}_i^x(t) \hat{\sigma}_j^y(t') \big>,
\end{equation}
and so on.
All expectation values and correlation functions at all combinations of times can be generated in this way, so in short, \textit{if two systems have the same generating functional, then their dynamics are identical}.

We begin by representing the generating functional as a Keldysh path integral, using $\hat{\sigma}^x$ eigenstates $| \sigma \rangle$ as the basis:
\begin{equation} \label{eq:proof_initial_path_integral}
\begin{aligned}
\mathcal{Z}[\xi] &\sim \sum_{\{ \sigma^+ \}, \{ \sigma^- \}} \left( \prod_t \big< \sigma^+(t + \Delta t) \big| e^{-i H(t) \Delta t + i \sum_i \vec{\xi}_i^+(t) \cdot \vec{\sigma}_i \Delta t} \big| \sigma^+(t) \big> \right) \big< \sigma^+(0) \big| \rho(0) \big| \sigma^-(0) \big> \\
&\qquad \qquad \qquad \qquad \qquad \qquad \qquad \qquad \qquad \qquad \cdot \left( \prod_t \big< \sigma^-(t) \big| e^{i H(t) \Delta t - i \sum_i \vec{\xi}_i^- \cdot \vec{\sigma}_i \Delta t} \big| \sigma^-(t + \Delta t) \big> \right).
\end{aligned}
\end{equation}
Here $\sigma^{\pm}$ is short-hand for the set $\{ \sigma_i^{\pm}(t) \}$, with $i \in \{1, \cdots, N\}$, $t \in \{0, \Delta t, \cdots, T - \Delta t, T\}$, and $\sigma_i^{\pm}(t) \in \{+1, -1\}$.
The sum over $\{ \sigma^+ \}, \{ \sigma^- \}$ denotes a nested sum over all possible values of $\sigma_i^{\pm}(t)$ for all $i$ and $t$ (with the boundary condition that $\sigma_i^+(T) = \sigma_i^-(T)$).
Similarly, the products are over all $t$. 
Lastly, the time-step $\Delta t$ should be interpreted as infinitesimally small.

Write the Hamiltonian as $H(t) = H_{\textrm{CA}}(t) + V_{\textrm{TI}}(t)$ (with CA denoting ``conventional annealing'' and TI denoting ``transverse interactions''), where
\begin{equation} \label{eq:proof_Hamiltonian_separation}
H_{\textrm{CA}}(t) \equiv s(t) \lambda(t) H_0 - (1 - s(t)) \sum_{i=1}^N \hat{\sigma}_i^x, \qquad V_{\textrm{TI}}(t) \equiv \frac{s(t) (1 - \lambda(t))}{N} \sum_{i,j=1}^N \hat{\sigma}_i^x \hat{\sigma}_j^x.
\end{equation}
For brevity, we also denote $F^{\pm}(t) \equiv \sum_i \vec{\xi}_i^{\pm}(t) \cdot \vec{\sigma}_i$.
Since $\Delta t$ is infinitesimal, the evolution operator $e^{-i H(t) \Delta t}$ factors:
\begin{equation} \label{eq:proof_short_evolution_factoring}
\begin{aligned}
\big< \sigma^+(t + \Delta t) \big| e^{-i [H(t) - F^+(t)] \Delta t} \big| \sigma^+(t) \big> &\sim \big< \sigma^+(t + \Delta t) \big| e^{-i [H_{\textrm{CA}}(t) - F^+(t)] \Delta t} e^{-i V_{\textrm{TI}}(t) \Delta t} \big| \sigma^+(t) \big> \\
&= e^{-i s(t) (1 - \lambda(t)) N^{-1} \sum_{ij} \sigma_i^+(t) \sigma_j^+(t) \Delta t} \big< \sigma^+(t + \Delta t) \big| e^{-i [H_{\textrm{CA}}(t) - F^+(t)] \Delta t} \big| \sigma^+(t) \big>,
\end{aligned}
\end{equation}
where the lower line uses that $| \sigma^+(t) \rangle$ is a $\hat{\sigma}^x$ eigenstate and thus $V_{\textrm{TI}}(t)$ is diagonal.
An analogous expression holds for the other set of matrix elements in Eq.~\eqref{eq:proof_initial_path_integral}.
Thus the generating functional becomes
\begin{equation} \label{eq:proof_path_integral_factored}
\begin{aligned}
\mathcal{Z}[\xi] &\sim \sum_{\{ \sigma^+ \}, \{ \sigma^- \}} e^{-i \sum_t s(t) (1 - \lambda(t)) N^{-1} \sum_{ij} [\sigma_i^+(t) \sigma_j^+(t) - \sigma_i^-(t) \sigma_j^-(t)] \Delta t} \\
&\qquad \qquad \qquad \cdot \left( \prod_t \big< \sigma^+(t + \Delta t) \big| e^{-i [H_{\textrm{CA}}(t) - F^+(t)] \Delta t} \big| \sigma^+(t) \big> \right) \big< \sigma^+(0) \big| \rho(0) \big| \sigma^-(0) \big> \\
&\qquad \qquad \qquad \qquad \qquad \qquad \qquad \qquad \qquad \qquad \qquad \cdot \left( \prod_t \big< \sigma^-(t) \big| e^{i [H_{\textrm{CA}}(t) - F^-(t)] \Delta t} \big| \sigma^-(t + \Delta t) \big> \right).
\end{aligned}
\end{equation}
Note that $\sum_{ij} \sigma_i^{\pm}(t) \sigma_j^{\pm}(t) = (\sum_i \sigma_i^{\pm}(t))^2$.
Thus if we introduce integration variables $m^{\pm}(t)$ alongside $\delta$-functions fixing $m^{\pm}(t) = N^{-1} \sum_i \sigma_i^{\pm}(t)$, we have that
\begin{equation} \label{eq:proof_path_integral_with_magnetization}
\begin{aligned}
\mathcal{Z}[\xi] &\sim \int \prod_t \textrm{d}m^+(t) \textrm{d}m^-(t) e^{-i N \sum_t s(t) (1 - \lambda(t)) [m^+(t)^2 - m^-(t)^2] \Delta t} \\
&\qquad \cdot \sum_{\{ \sigma^+ \}, \{ \sigma^- \}} \prod_t \delta \Big( m^+(t) - N^{-1} \sum_i \sigma_i^+(t) \Big) \delta \Big( m^-(t) - N^{-1} \sum_i \sigma_i^-(t) \Big) \\
&\qquad \qquad \qquad \cdot \left( \prod_t \big< \sigma^+(t + \Delta t) \big| e^{-i [H_{\textrm{CA}}(t) - F^+(t)] \Delta t} \big| \sigma^+(t) \big> \right) \big< \sigma^+(0) \big| \rho(0) \big| \sigma^-(0) \big> \\
&\qquad \qquad \qquad \qquad \qquad \qquad \qquad \qquad \qquad \qquad \qquad \cdot \left( \prod_t \big< \sigma^-(t) \big| e^{i [H_{\textrm{CA}}(t) - F^-(t)] \Delta t} \big| \sigma^-(t + \Delta t) \big> \right).
\end{aligned}
\end{equation}
Lastly, use the Fourier representation of the $\delta$-functions, i.e., $\delta(x) \propto \int \textrm{d}\Gamma e^{i \Gamma x}$, to write
\begin{equation} \label{eq:proof_path_integral_both_variables}
\begin{aligned}
\mathcal{Z}[\xi] &\sim \int \prod_t \textrm{d}m^+(t) \textrm{d}m^-(t) \textrm{d}\Gamma^+(t) \textrm{d}\Gamma^-(t) e^{-i N \sum_t \big[ s(t) (1 - \lambda(t)) [m^+(t)^2 - m^-(t)^2] - \Gamma^+(t) m^+(t) + \Gamma^-(t) m^-(t) \big]} \mathcal{Z}_{\textrm{SCE}}[\xi],
\end{aligned}
\end{equation}
where
\begin{equation} \label{eq:proof_self_consistent_density_matrix}
\begin{aligned}
\mathcal{Z}_{\textrm{SCE}}[\xi] &\equiv \sum_{\{ \sigma^+ \}, \{ \sigma^- \}} \left( \prod_t \big< \sigma^+(t + \Delta t) \big| e^{-i [H_{\textrm{CA}}(t) - F^+(t) + \Gamma^+(t) \sum_i \hat{\sigma}_i^x] \Delta t} \big| \sigma^+(t) \big> \right) \big< \sigma^+(0) \big| \rho(0) \big| \sigma^-(0) \big> \\
&\qquad \qquad \qquad \qquad \qquad \qquad \qquad \qquad \qquad \qquad \cdot \left( \prod_t \big< \sigma^-(t) \big| e^{i [H_{\textrm{CA}}(t) - F^-(t) + \Gamma^-(t) \sum_i \hat{\sigma}_i^x] \Delta t} \big| \sigma^-(t + \Delta t) \big> \right).
\end{aligned}
\end{equation}

The integrand in Eq.~\eqref{eq:proof_path_integral_both_variables} is of the form\footnote{Arguing that $\mathcal{Z}_{\textrm{SCE}}$ scales exponentially with $N$ is somewhat subtle and admittedly not rigorous. Note that, except for factors of $i$, $\mathcal{Z}_{\textrm{SCE}}$ is analogous to a many-body partition function: it is the sum of the exponential of a Hamiltonian over all possible states of $O(N)$-many degrees of freedom (the Ising spins $\sigma_i^{\pm}(t)$). Since the Hamiltonian is extensive, i.e., $O(N)$, one would expect that the ``free energy'' $\log{\mathcal{Z}_{\textrm{SCE}}}$ is also extensive, equivalently that $\mathcal{Z}_{\textrm{SCE}}$ scales exponentially with $N$.} $e^{-iN \cdots}$.
Thus at large $N$, we can evaluate the integrals over $m^{\pm}$ and $\Gamma^{\pm}$ by saddle-point approximation: set $m^{\pm}$ and $\Gamma^{\pm}$ to the values at which the exponent of the integrand is stationary, i.e., the derivatives of the exponent are zero.
This gives the equations
\begin{equation} \label{eq:proof_self_consistent_equations_initial}
\Gamma^{\pm}(t) = 2 s(t) \big( 1 - \lambda(t) \big) m^{\pm}(t), \qquad m^{\pm}(t) = \pm \frac{i}{N} \frac{\partial \log{\mathcal{Z}_{\textrm{SCE}}}[\xi]}{\partial \Gamma^{\pm}(t)}.
\end{equation}
A self-consistent solution has $m^+(t) = m^-(t)$ and $\Gamma^+(t) = \Gamma^-(t)$.
This is clear for the left-hand equation in Eq.~\eqref{eq:proof_self_consistent_equations_initial}.
As for the right-hand equation, note that if we set $\Gamma^+(t) = \Gamma^-(t)$, then $\mathcal{Z}_{\textrm{SCE}}$ becomes the generating functional for a system evolving under Hamiltonian $H_{\textrm{CA}}(t) + \Gamma(t) \sum_i \hat{\sigma}_i^x$.
For one thing, this means that $\mathcal{Z}_{\textrm{SCE}}[0] = 1$ (and we are only interested in the generating functional near $\xi = 0$ since that is all that we need to calculate observables).
Furthermore, by analogy with Eq.~\eqref{eq:proof_generating_functional_derivatives},
\begin{equation} \label{eq:proof_self_consistent_field_derivatives}
i \frac{\delta \mathcal{Z}_{\textrm{SCE}}[0]}{\delta \Gamma^+(t)} = -i \frac{\delta \mathcal{Z}_{\textrm{SCE}}[0]}{\delta \Gamma^-(t)} = \sum_i \big< \hat{\sigma}_i^x(t) \big>_{\textrm{SCE}}.
\end{equation}
Thus Eq.~\eqref{eq:proof_self_consistent_equations_initial} amounts to
\begin{equation} \label{eq:proof_self_consistent_equations}
\Gamma(t) = 2 s(t) \big( 1 - \lambda(t) \big) m(t), \qquad m(t) = \frac{1}{N} \sum_i \big< \hat{\sigma}_i^x(t) \big>_{\textrm{SCE}}.
\end{equation}

When $m^+(t) = m^-(t)$ and $\Gamma^+(t) = \Gamma^-(t)$, the exponential in Eq.~\eqref{eq:proof_path_integral_both_variables} becomes unity and we are left with
\begin{equation} \label{eq:proof_dynamics_equivalence}
\mathcal{Z}[\xi] \sim \mathcal{Z}_{\textrm{SCE}}[\xi].
\end{equation}
As discussed above, this means that the systems corresponding to these two generating functionals have equivalent dynamics.
The left-hand generating functional is that of the original Hamiltonian with transverse interactions (Eq.~\eqref{eq:proof_original_Hamiltonian}).
The right-hand generating functional, meanwhile, is seen to be precisely that of SCE: the system evolves under Hamiltonian $H_{\textrm{CA}}(t) + \Gamma(t) \sum_i \hat{\sigma}_i^x$, which does not have the transverse interactions but instead an additional transverse field, such that $\Gamma(t)$ is given by $2 s(t) (1 - \lambda(t))$ times the average $x$-magnetization at that instant.
This establishes the equivalence between ED and SCE in the large-$N$ limit.

\section{Alternate interpolation schemes for SCD} \label{sec:alternate_interpolation}

In the main text, we consider the simplest scheme for approximating the SCE protocol by one with a finite waiting time $w$: take $\Gamma_{\textrm{SCD}}(t)$ to be piecewise-constant, only updating at times which are multiples of $w$.
The resulting $\Gamma_{\textrm{SCD}}(t)$ is of course discontinuous, which may not be feasible on certain platforms.
Thus here we consider the next level of refinement in the self-consistent discrete protocol, and take $\Gamma_{\textrm{SCD}}(t)$ to linearly interpolate between the magnetization at multiples of $w$.
Written out, for $t \in [nw, (n+1)w)$, we set
\begin{equation} \label{eq:self_consistent_discrete_alternate}
\Gamma_{\textrm{SCD}}(t) = \frac{1}{N} \sum_{i=1}^N \big< \hat{\sigma}_i^x(nw) \big> + \frac{t - nw}{w} \cdot \frac{1}{N} \sum_{i=1}^N \Big( \big< \hat{\sigma}_i^x((n+1)w) - \big< \hat{\sigma}_i^x(nw) \big> \Big).
\end{equation}
This choice for $\Gamma_{\textrm{SCD}}(t)$ is at least continuous, although it has a kink at each multiple of $w$.
One could of course generalize Eq.~\eqref{eq:self_consistent_discrete_alternate} to higher-order interpolations that maintain the continuity of derivatives of $\Gamma_{\textrm{SCD}}(t)$, but we do not consider this here.

An issue with Eq.~\eqref{eq:self_consistent_discrete_alternate} is that it does not specify the field for $t$ beyond the last time at which the magnetization was measured.
In other words, suppose that we have determined the magnetization up to time $n_m w$.
We need to define $\Gamma_{\textrm{SCD}}(t)$ up to time $(n_m + 1)w$ before we can determine the magnetization at $(n_m + 1)w$.
As the simplest choice, we temporarily set $\Gamma_{\textrm{SCD}}(t) = N^{-1} \sum_i \langle \hat{\sigma}_i^x(n_m w) \rangle$ for $t \in [n_m w, (n_m + 1) w)$, then measure the magnetization at $(n_m + 1) w$ and use Eq.~\eqref{eq:self_consistent_discrete_alternate} for $t \in [n_m w, (n_m + 1) w)$ from then on.
This means that $\Gamma_{\textrm{SCD}}(nw)$ will not quite match $N^{-1} \sum_i \langle \hat{\sigma}_i^x(nw) \rangle$, but importantly, this discrepancy vanishes in the limit $w \rightarrow 0$.
Thus SCD can still be made to agree with SCE arbitrarily well by taking $w$ sufficiently small.

\begin{figure}
\centering
\includegraphics[width=0.5\linewidth]{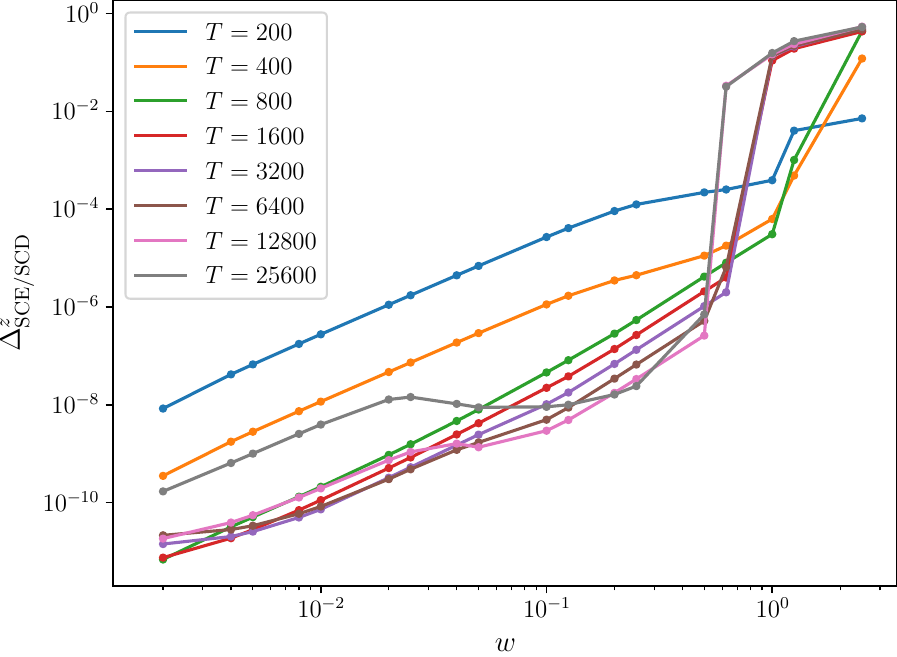}
\caption{Error between SCE and the linear-interpolation variant of SCD (Eq.~\eqref{eq:self_consistent_discrete_alternate}), as a function of the waiting time $w$ for various values of the runtime $T$. System size is $N = 100$, and the problem Hamiltonian has $p = 3$ and $h = 1$.}
\label{fig:interpol}
\end{figure}

We plot the error between SCE and this variant of SCD in Fig.~\ref{fig:interpol}, analogous to the right panel of Fig.~\ref{fig:SCD0}.
Qualitatively, the trend is the same: the error is extremely small at small $w$, close to maximal at large $w$, and the crossover between the two sharpens as the runtime $T$ increases.
The crossover also drifts towards smaller $w$ as $T$ increases, suggesting that SCD will still break down at sufficiently large $T$ for any non-zero $w$.
That said, the error here is much smaller in absolute terms, as makes sense since the linear interpolation provides a much better approximation to the trajectory of the magnetization.
Presumably higher-order interpolations would have even smaller errors.
Thus refining the interpolation scheme provides another means of improving the performance of the self-consistent protocol, separate from simply decreasing $w$.

\section{Self-consistent protocol in the presence of phase transitions} \label{sec:phase_transitions}

\begin{figure}
\centering
\includegraphics[width=0.5\linewidth]{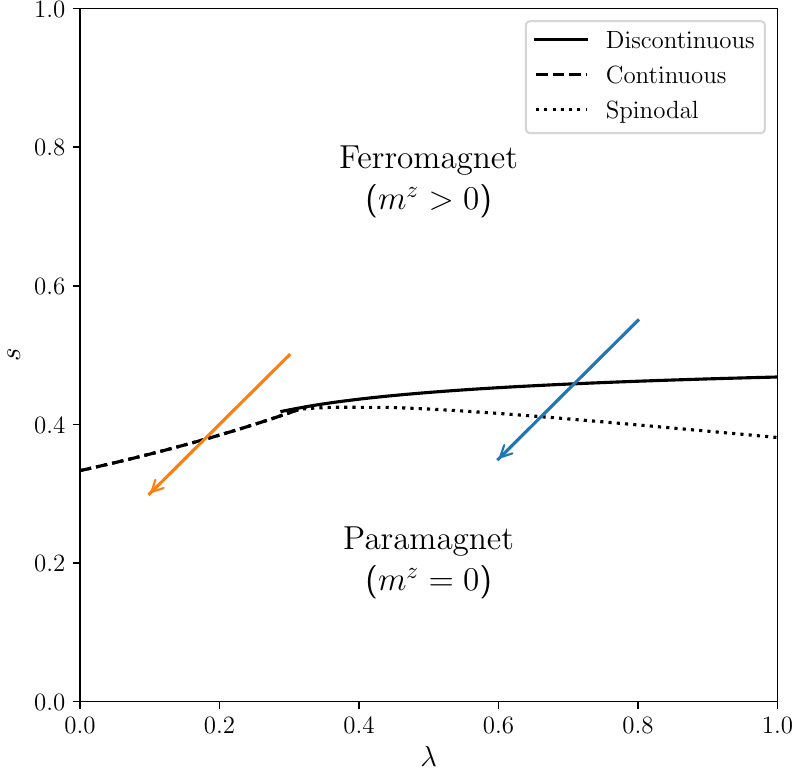}
\caption{Phase diagram (black lines) of the $p$-spin model with a transverse field and transverse interactions (i.e., Eq.~\eqref{eq:ED_hamiltonian} using Eq.~\eqref{eq:p_spin_model} for $H_0$), with $p = 5$ and $h = 0$. We consider protocols that follow the paths in orange and blue --- the orange path starts at $(0.3, 0.5)$ and ends at $(0.1, 0.3$), while the blue path starts at $(0.8, 0.55)$ and ends at $(0.6, 0.35)$.}
\label{fig:phase_diagram_p5}
\end{figure}

Here we give results analogous to those in the main text for the $p$-spin model with $p = 5$ and $h = 0$.
Unlike the example used for the main text, this model has a non-trivial phase diagram as a function of $s$ and $\lambda$, shown in Fig.~\ref{fig:phase_diagram_p5}.
The purpose of this appendix is to show how the errors in the self-consistent approach behave when the protocol passes through phase transitions.
While the details of the trends show a number of differences compared to the transition-free protocol, we find that the overall qualitative conclusions are quite similar. 

Note that this does \textit{not} mean that the self-consistent protocol is equally effective in the presence of phase transitions --- as usual, the runtime required to remain adiabatic is much longer when crossing transitions than when not.
Instead, it is the ability of the self-consistent protocol to emulate transverse interactions that is qualitatively insensitive to transitions, at least in terms of the rough dependence on parameters.

For completeness, let us first mention how to determine the phase diagram.
While the derivation of this procedure is somewhat involved (see Refs.~\cite{QuantumAnnealingAntiferromagnetic2012seki,ManybodyTransverseInteractions2012seoane}), the procedure itself is straightforward to state.
First express the Hamiltonian in terms of the total spin operators $\hat{S}^{\alpha} \equiv \sum_i \hat{\sigma}_i^{\alpha}$:
\begin{equation} \label{eq:Hamiltonian_total_spin}
H = -N s \lambda \big( N^{-1} \hat{S}^z \big)^p - s \lambda h \hat{S}^z + N s (1 - \lambda) \big( N^{-1} \hat{S}^x \big)^2 - (1 - s) \hat{S}^x.
\end{equation}
Then replace $\hat{S}^{\alpha}/N$ by a \textit{classical} unit vector $\hat{m} \equiv (m^x, m^y, m^z)$, giving a classical energy landscape $\epsilon(\hat{m})$ (we divide by an overall factor of $N$ for convenience):
\begin{equation} \label{eq:classical_energy_landscape}
\epsilon(\hat{m}) \equiv -s \lambda (m^z)^p - s \lambda h m^z + s (1 - \lambda) (m^x)^2 - (1 - s) m^x.
\end{equation}
The ground-state energy density is simply the minimum of $\epsilon(\hat{m})$, and the ground-state magnetization is the location of that minimum.
As usual, points in parameter space where the minimizing value of $\hat{m}$ jumps discontinuously correspond to discontinuous phase transitions, and points where it changes continuously but non-differentiably correspond to continuous transitions.

The resulting phase diagram for $p = 5$ and $h = 0$, shown in Fig.~\ref{fig:phase_diagram_p5}, has both continuous and discontinuous phase transitions.
It is much more advantageous for QA to pass through the continuous phase boundary, but again, the focus of this manuscript is rather on how to emulate transverse interactions regardless of the path in parameter space.
Thus we consider both a path (orange) passing through the continuous phase boundary and a path (blue) passing through the discontinuous boundary (note that the blue path also passes through the spinodal, i.e., the line along which the ferromagnetic minimum in $\epsilon(\hat{m})$ becomes locally unstable).
As in the main text, we compute the error between subsequent versions of the self-consistent protocol --- ED vs.\ SCE vs.\ SCD vs.\ SCM --- still using Eq.~\eqref{eq:error_figure_merit} as the figure of merit.

\begin{figure}
\centering
\includegraphics[width=0.48\linewidth]{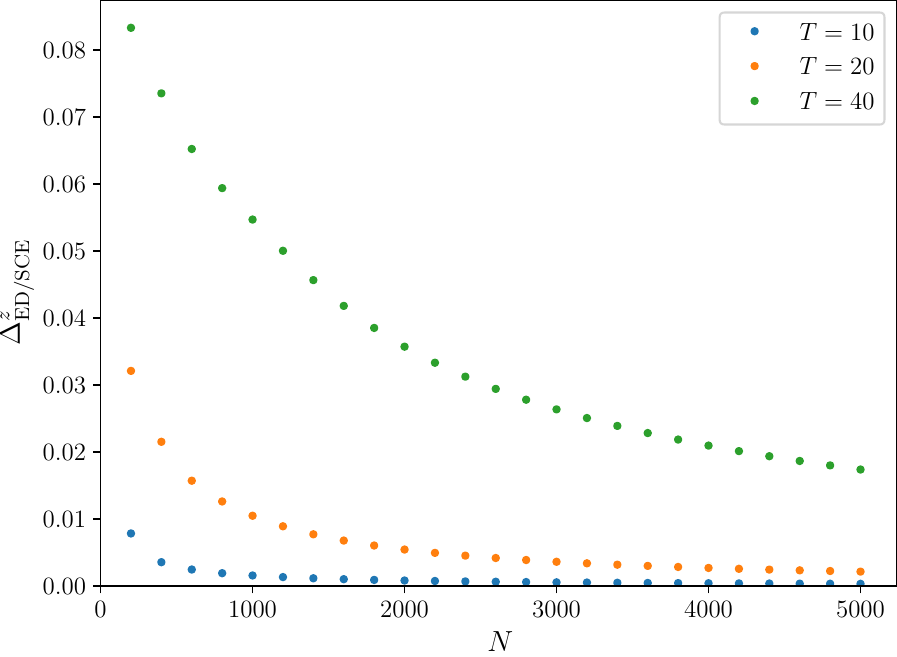}
\quad
\includegraphics[width=0.48\linewidth]{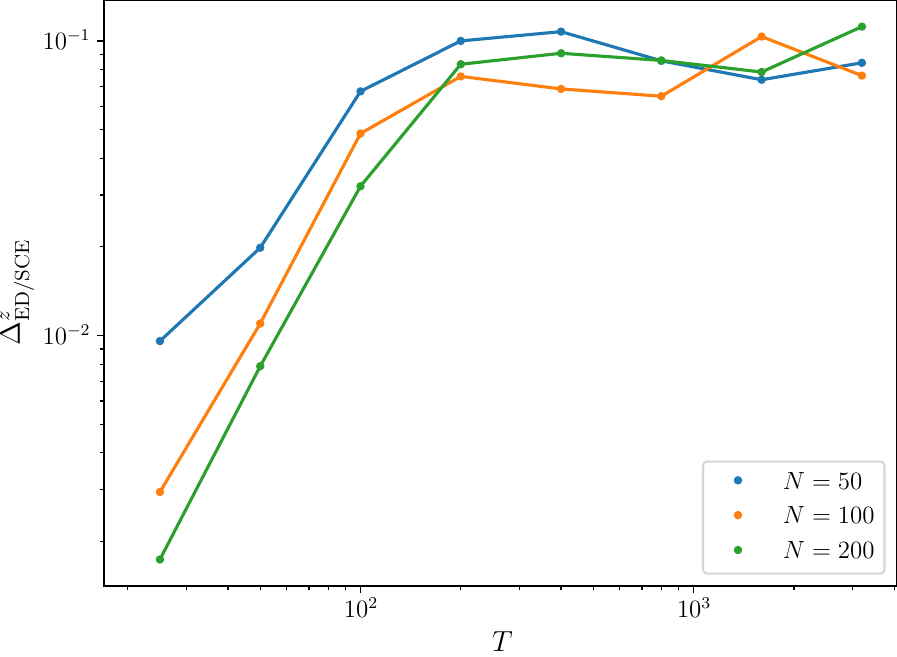} \\
\includegraphics[width=0.48\linewidth]{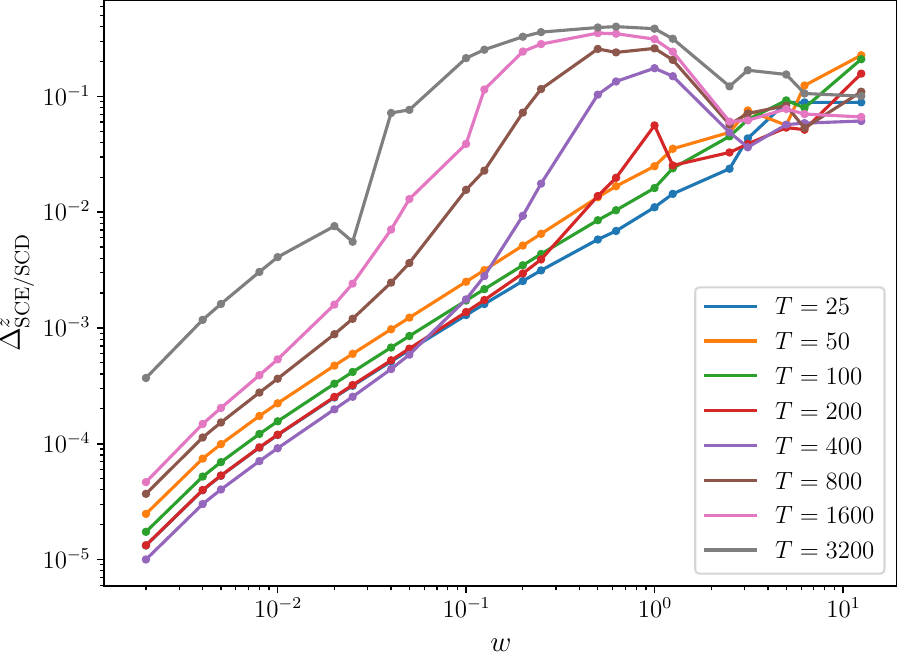}
\quad
\includegraphics[width=0.48\linewidth]{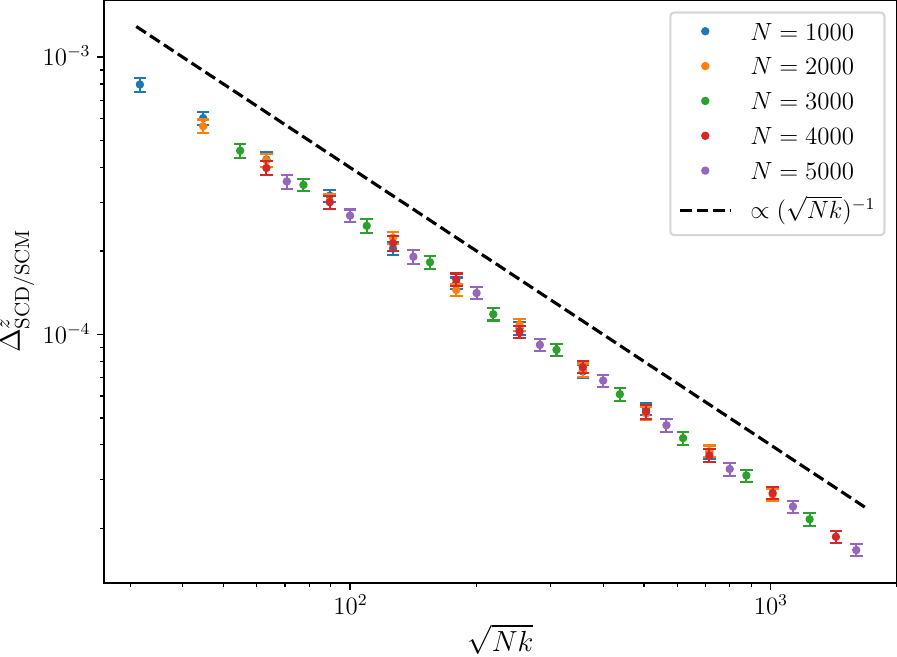}
\caption{Errors between protocols following the blue path in Fig.~\ref{fig:phase_diagram_p5} (crossing the discontinuous boundary and spinodal). Top-left compares ED to SCE as a function of $N$, top-right compares ED to SCE as a function of $T$, bottom-left compares SCE to SCD as a function of $w$ and $T$ (at $N = 100$), and bottom-right compares SCD to SCM as a function of $\sqrt{Nk}$ (at $T=25$ and $w=0.01$).}
\label{fig:sp}
\end{figure}

\begin{figure}
\centering
\includegraphics[width=0.48\linewidth]{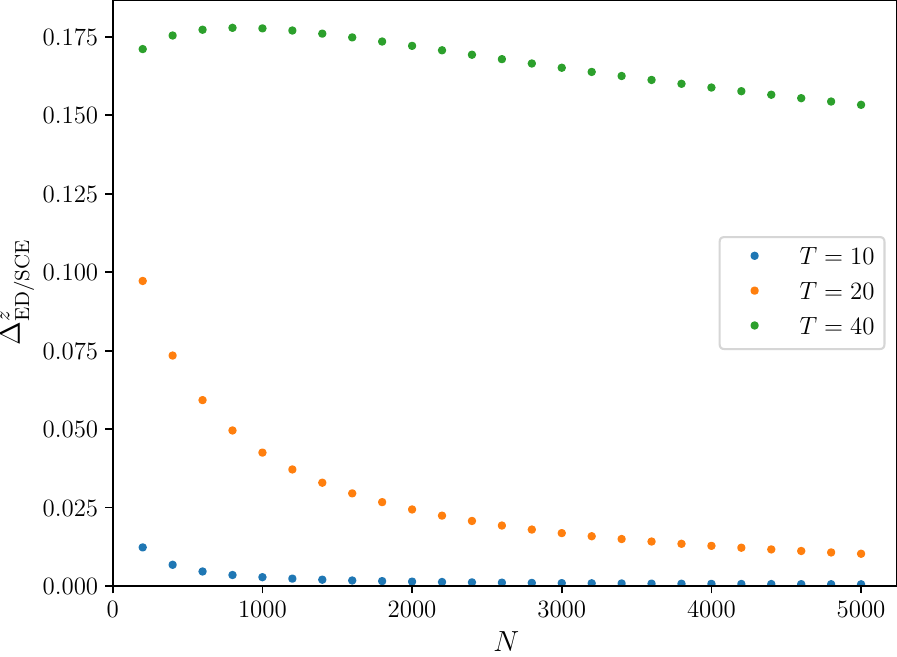}
\quad
\includegraphics[width=0.48\linewidth]{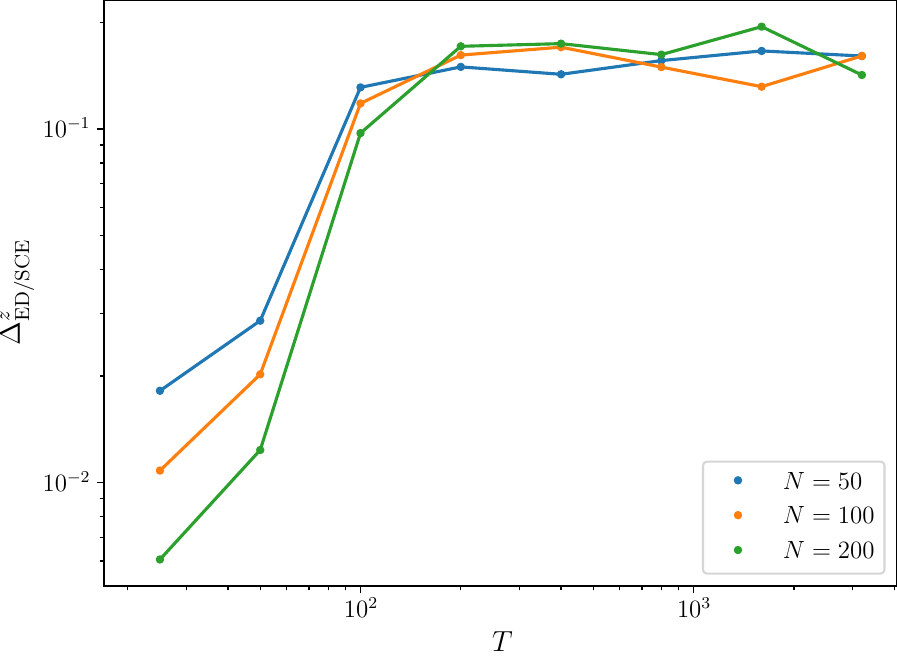}\\
\includegraphics[width=0.48\linewidth]{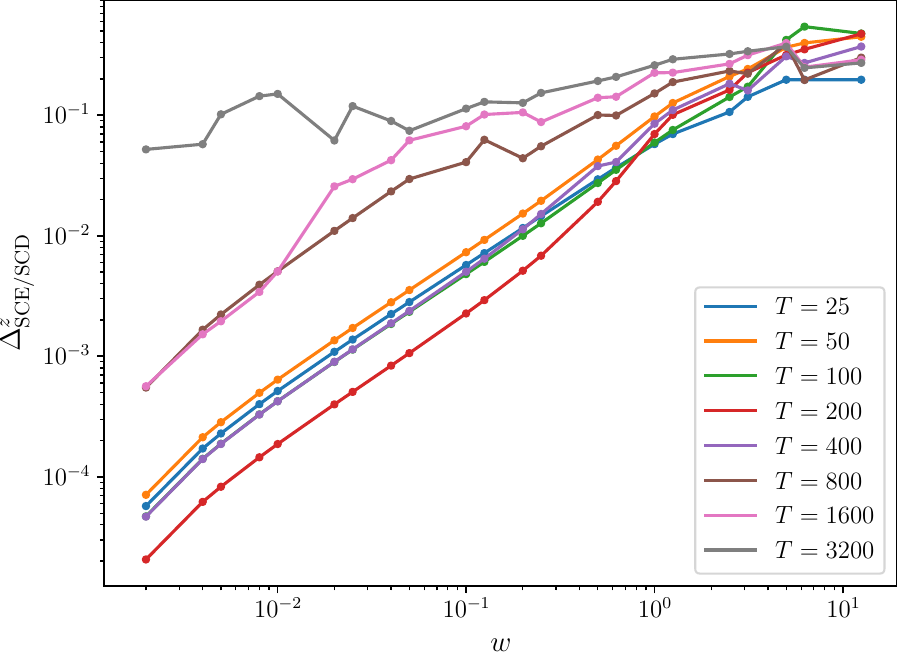}
\quad
\includegraphics[width=0.48\linewidth]{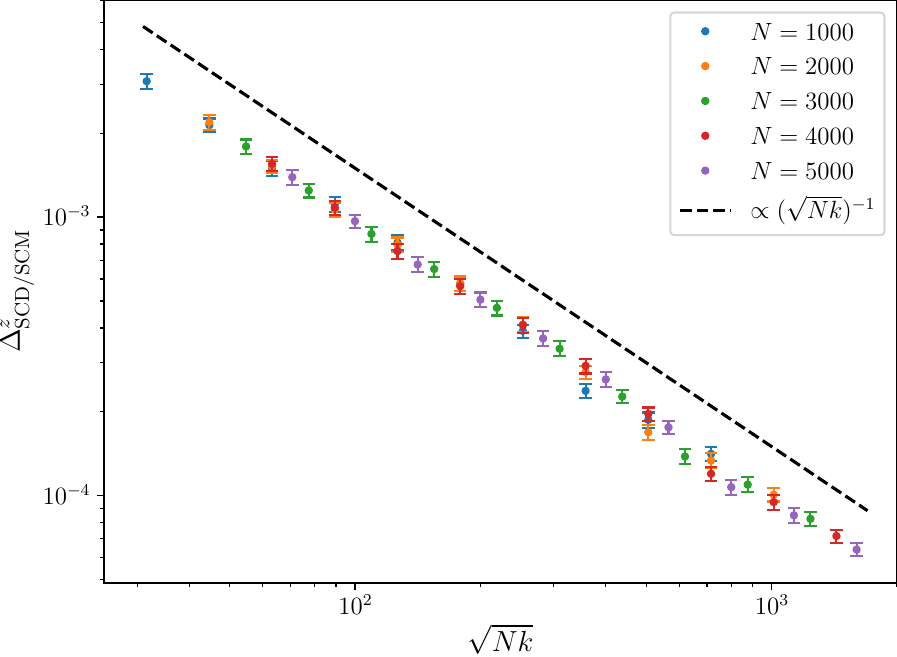}
\caption{Errors between protocols following the orange path in Fig.~\ref{fig:phase_diagram_p5} (crossing the continuous boundary). Top-left compares ED to SCE as a function of $N$, top-right compares ED to SCE as a function of $T$, bottom-left compares SCE to SCD as a function of $w$ and $T$ (at $N = 100$), and bottom-right compares SCD to SCM as a function of $\sqrt{Nk}$ (at $T=25$ and $w=0.01$).}
\label{fig:2nd order}
\end{figure}

Fig.~\ref{fig:sp} shows the errors when crossing the discontinuous phase boundary, and Fig.~\ref{fig:2nd order} shows the errors when crossing the continuous boundary.
The top two panels in each figure compare ED to SCE.
As a function of $N$ at fixed $T$, the error indeed decreases to zero, although it is interesting to note that the convergence is much slower when passing through the continuous phase transition (it is even hard to argue from the data alone that $\Delta_{\textrm{ED/SCE}}^z$ is tending to zero for $T = 40$, but the derivation in Appendix~\ref{sec:appendixb} leads us to believe that it ultimately will for sufficiently large $N$).
As a function of $T$, we do find one noticeable difference from the example in the main text --- here the error increases as a function of $T$ in both cases.
It does still saturate at large $T$, although the asymptotic value is much larger than that in the main text (roughly $10^{-1}$ versus $10^{-3}$).
Despite these differences, however, the overall implication for QA is the same: there is a limit to how well the self-consistent protocol can perform as a function of $T$.

The bottom-left panel in each figure compares SCE to SCD.
As a function of $w$ at fixed $T$, the error goes to zero as $w \rightarrow 0$ and is (roughly) monotonic as expected.
As a function of $T$, while there is perhaps a slight vestige of the instability and crossover behavior seen in the main text, the overall trend is for the error to simply increase with $T$ for all $w$.
Yet once again, the different shape of the curves does not affect the broader message: SCD will ultimately break down at sufficiently large $T$ for any non-zero $w$.

Lastly, the bottom-right panel in each figure compares SCD to SCM.
The error is plotted as a function of $k$ for multiple values of $N$, but with the $x$-axis rescaled to be $\sqrt{Nk}$.
The fact that all points collapse extremely well onto a single straight line demonstrates that the error still scales as $1/\sqrt{Nk}$, exactly as in the main text.
In particular, we find the same trade-off in which less measurements are needed at larger system sizes.

To summarize, the conclusion is largely the same: even in the presence of phase transitions, the error introduced at each stage of approximation can be made sufficiently small by tuning the corresponding parameter ($N$, $w$, $k$), but with the important caveat that the performance of the self-consistent protocol will ultimately degrade at large $T$.

\end{document}